\documentclass[12pt,preprint]{aastex}
\usepackage{epsfig}

\shorttitle{Multi-threaded prominence}
\shortauthors{M. Luna et al.}

\begin{document}

\title{Formation and evolution of a multi-threaded prominence}

\author{M. Luna\altaffilmark{1}, J. T.
Karpen\altaffilmark{2}, and C. R. DeVore\altaffilmark{3}}

\altaffiltext{1}{CRESST and Space Weather Laboratory NASA/GSFC, Greenbelt, MD 20771, USA}
\altaffiltext{2}{NASA/GSFC, Greenbelt, MD 20771, USA}
\altaffiltext{3}{Naval Research Laboratory, Washington, DC 20375, USA}

\begin{abstract}
We investigate the process of formation and subsequent evolution of prominence plasma in a filament channel and its overlying arcade. We construct a three-dimensional time-dependent model of an intermediate quiescent prominence suitable to be compared with observations. We combine the magnetic field structure with one-dimensional independent simulations of many flux tubes, of a three-dimensional sheared double arcade, in which the thermal nonequilibrium process governs the plasma evolution. We have found that the condensations in the corona can be divided into two populations: threads and blobs. Threads are massive condensations that linger in the field line dips. Blobs are ubiquitous small condensations that are produced throughout the filament and overlying arcade magnetic structure, and rapidly fall to the chromosphere. The threads are the principal contributors to the total mass, whereas the blob contribution is small. The total prominence mass is in agreement with observations, assuming reasonable filling factors of order $0.001$ and a fixed number of threads. The motion of the threads is basically horizontal, while blobs move in all directions along the field. The peak velocities for both populations are comparable, but there is a weak tendency for the velocity to increase with the inclination, and the blobs with nearly vertical motion have the largest velocities. We have generated synthetic images of the whole structure in an H$\alpha$ proxy and in two EUV channels of the AIA instrument aboard SDO, thus showing the plasma at cool, warm, and hot temperatures. The predicted differential emission measure of our system agrees very well with observations in the temperature range $\log\,T = 4.6-5.7$. We conclude that the sheared-arcade magnetic structure and plasma behavior driven by thermal nonequilibrium fit well the abundant observational evidence for typical intermediate prominences.
\end{abstract}

\section{Introduction}

Quiescent solar prominences have been studied for a long time, and significant progress in determining their nature has been made. We can say with certainty that they are composed of cold plasma, and it is generally accepted that the magnetic field supports these dense structures in the tenuous corona. However, there are still important unknowns about their structure, formation, and evolution. High-resolution observations from space and from the ground --– e.g., from SOHO, TRACE, SVST, VAULT, Hinode, and SDO –-- show a complex fine structure in space and time. This fine structure consists of moving blobs and bundles of horizontal threads with counterstreaming motions \citep{engvold1976,zirker1998,lin2003,vourlidas2003,lin2005,berger2008}. The quiescent prominences are immersed in structures even larger than themselves. Prominences form above a polarity inversion line \citep{babcock1955,smith1967,mcinstosh1972} in filament channels \citep{martin1998,gaizauskas2001}, which are regions of sheared magnetic field \citep{hagyard1984,moore1987,Venkatakrishnan1989}.  Magnetic loop arcades overlie filament channels, spanning high above them and rooted at both sides of the channel \citep{tandberg1995,martin1998}. Direct measurements of the magnetic fields have shed some light on the magnetic structure of quiescent prominences \citep{leroy1989,casini2003,Kuckein2009}. However, these measurements are focused on the cool prominence plasma alone and do not establish definitively the global structure of the magnetic field in the coronal portion of the filament channel. 

There are basically two models of the magnetic structure of filament channels: the sheared arcade and the flux rope \citep[see review by][]{mackay2010}. \citet{antiochos1994}, \cite{devore2000}, and \citet{aulanier2002} showed that the sheared arcade model exhibits most of the observed properties of the large-scale filament channel structure. The generated magnetic field structure has dipped and flattened field lines consistent with most prominence observations \citep{martin1998}. Many quiescent filaments are observed to grow by linkage between smaller segments \citep{malherbe1989,gaizauskas2001,schmieder2004,vanballegooijen2004}. \citet{devore2005} and \citet{aulanier2006} studied the interaction of two neighboring sheared bipoles with different combinations of polarity and chirality. They found that merging two adjacent sheared arcades with aligned axial field formed a long, stable, strongly sheared prominence. Assuming that the cool plasma resides in magnetic dips, the resulting structure reproduces the global appearance of a typical low-latitude intermediate quiescent filament \citep{mackay2010}. The present work goes beyond this idealized assumption to determine the origin and time-dependent evolution of the cool prominence plasma.

It is generally believed that the origin of the prominence plasma is the chromosphere \citep{pikelner1971,saito1973,zirker1994}, but the mechanism that puts the chromospheric material into the corona is under debate. One promising candidate, the thermal nonequilibrium process, has been studied extensively \citep{antiochos1991,antiochos1999,antiochos2000,karpen2001,karpen2003,karpen2005,karpen2006,karpen2008}, and yields quantitative predictions of prominence plasma directly comparable with observations. In this model, localized heating above the flux tube footpoints produces evaporation of the chromospheric plasma, which condenses in the coronal part of the tube. The localized heating near the chromosphere is consistent with several coronal-loop studies \citep{aschwanden2001,schmieder2004b}. To produce this instability the heating must be concentrated in a region of approximately one eighth of the tube length or less \citep{serio1981,mok1990,antiochos1991,dahlburg1998}. In our earlier investigations of the thermal nonequilibrium process, we found that large condensations form along long, low-lying magnetic field lines. In most cases the plasma undergoes repeated dynamic cycles of plasma condensation, streaming along the tube, and falling to the chromosphere. Small, short-lived knots are produced in coronal loops, possibly explaining coronal rain \citep{schrijver2001,muller2003,muller2004,muller2005}. In deeply dipped flux tubes, condensations form readily but the subsequent evolution depends on the tube slope. For slopes steeper than a critical value determined by the heating asymmetry and gravitational scale height, the condensations fall to the bottom of the dip and remain there, accreting mass continuously. 

In the present work, we built a comprehensive model of a multi-threaded prominence combining the results of the three-dimensional (3D) sheared double-arcade model of \citet{devore2005} for the magnetic structure with the thermal nonequilibrium process for the plasma evolution determined by many independent one-dimensional simulations. The resulting three-dimensional model allowed us to study formation and evolution of the plasma in a multi-stranded prominence. In \S \ref{model-sec} we describe the theoretical model used in the current work. The detailed results of the investigation are presented in \S \ref{results-sec}. The global appearance of the prominence in different spectral channels and the predicted differential emission measure are shown in \S \ref{obser-sec}. Finally, in \S \ref{discussion-sec} the results of this investigation are summarized and conclusions are drawn. In particular, we conclude that our three-dimensional model fits well the observational evidence.

\section{Theoretical model}\label{model-sec}
As in our earlier studies, we assume that the magnetic structure is static on the timescale of the plasma evolution \citep[see e.g.,][]{martin1998} and that the gas pressure of the plasma is negligible in contrast with the magnetic pressure. Additionally, in the solar corona the conductive heat flux is mainly constrained to follow the field lines. With these considerations each flux tube is a thermally isolated system where the plasma motions are along the magnetic field. The equations describing the plasma evolution are reduced to one-dimensional (1D) set of hydrodynamic equations for mass, momentum and energy conservation \citep[see, e.g.,][]{karpen2005} including the Raymond-Klimchuk radiative loss function \citep{klimchuk2001}, \citet{spitzer1962} thermal conductivity, and a prescribed volumetric heating. Hence the time-dependent plasma evolution along each static field line can be determined by solving numerically those equations.

\begin{figure}[!ht]
\centering
\includegraphics[width=15cm]{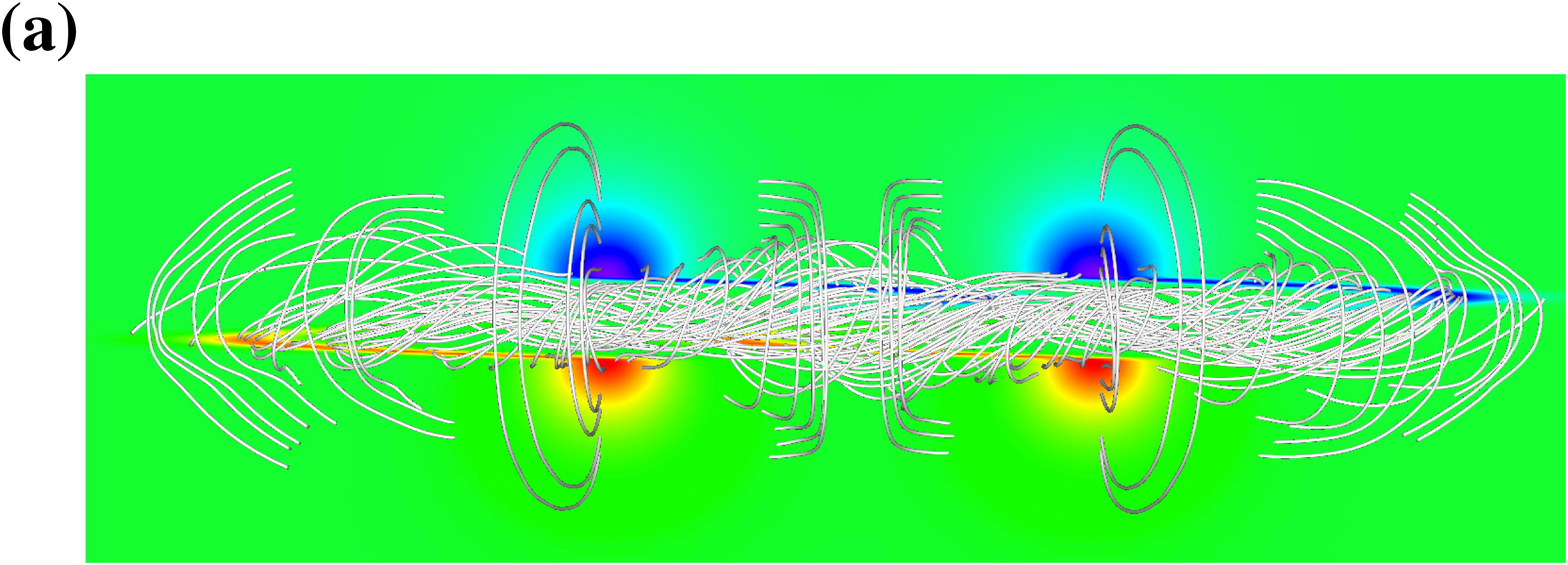}
\includegraphics[width=15cm]{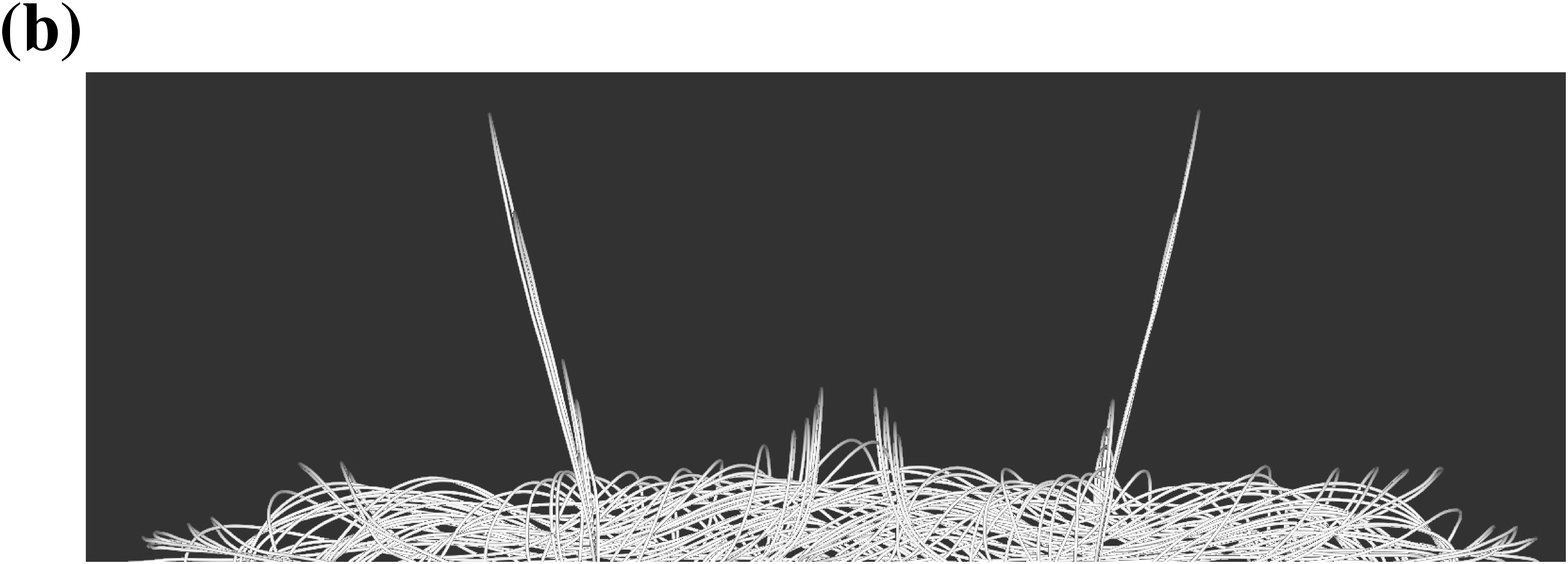}
\includegraphics[width=15cm]{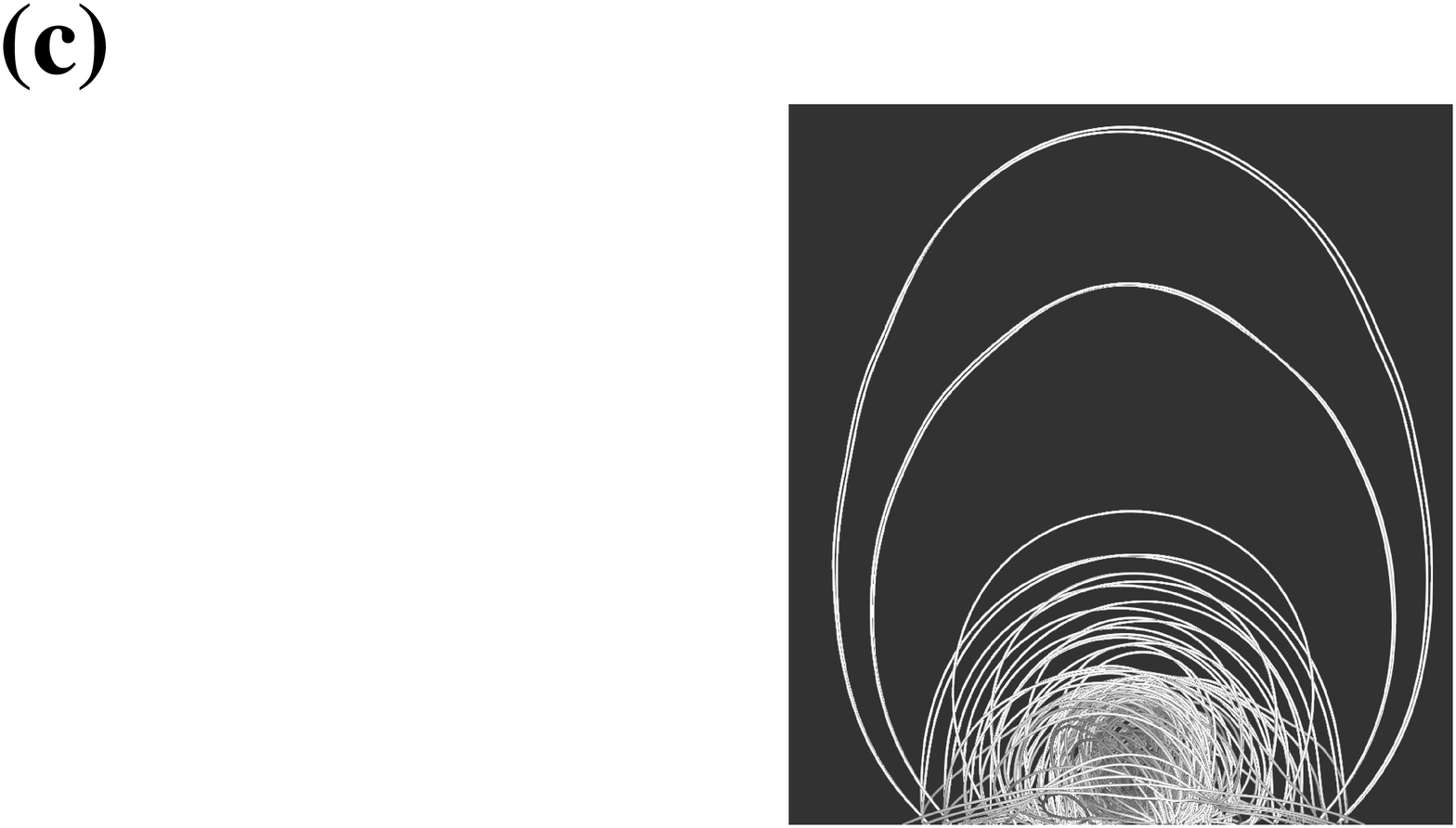}
\caption{Magnetic field structure of the sheared double arcade in Cartesian geometry. White lines are the magnetic field lines we have used in this work. In {\bf (a)} the magnetic structure is viewed from above. At the bottom of the corona the normal magnetic field is plotted (color scale). The two dipoles are clearly seen in this view. In {\bf (b)} the side-view orientation is shown. Finally, in {\bf (c)} the end-view orientation is plotted.}
\label{magnetic_structure}
\end{figure}

We use the magnetic field structure given by the sheared double-arcade model of \citet{devore2005} in Cartesian geometry. This magnetic field configuration is the result of shearing two identical dipoles with the same polarity, the same chirality, and with their axes aligned. The shear is produced by imposing differential footpoint motions over the photospheric plasma near the polarity inversion line (PIL). As a consequence, an inner bundle of sheared field lines is produced (see Fig. \ref{magnetic_structure}). These are long, low-lying field lines nearly parallel to the PIL, and many of them have dips. An outer bundle of un-sheared field lines comprises the overlying arcade with arched field lines perpendicular (or slightly skewed from perpendicular) to the PIL. In this model, the filament channel (FC) magnetic configuration is a natural consequence of the magnetic shear in a 3D structure. However, the exact origin of the magnetic shear is not important in our model. The shear could be the result of the partial emergence of twisted flux tubes from below, or direct subsurface motions near the PIL.

\begin{figure}[ht]
\centering
\includegraphics[width=12cm]{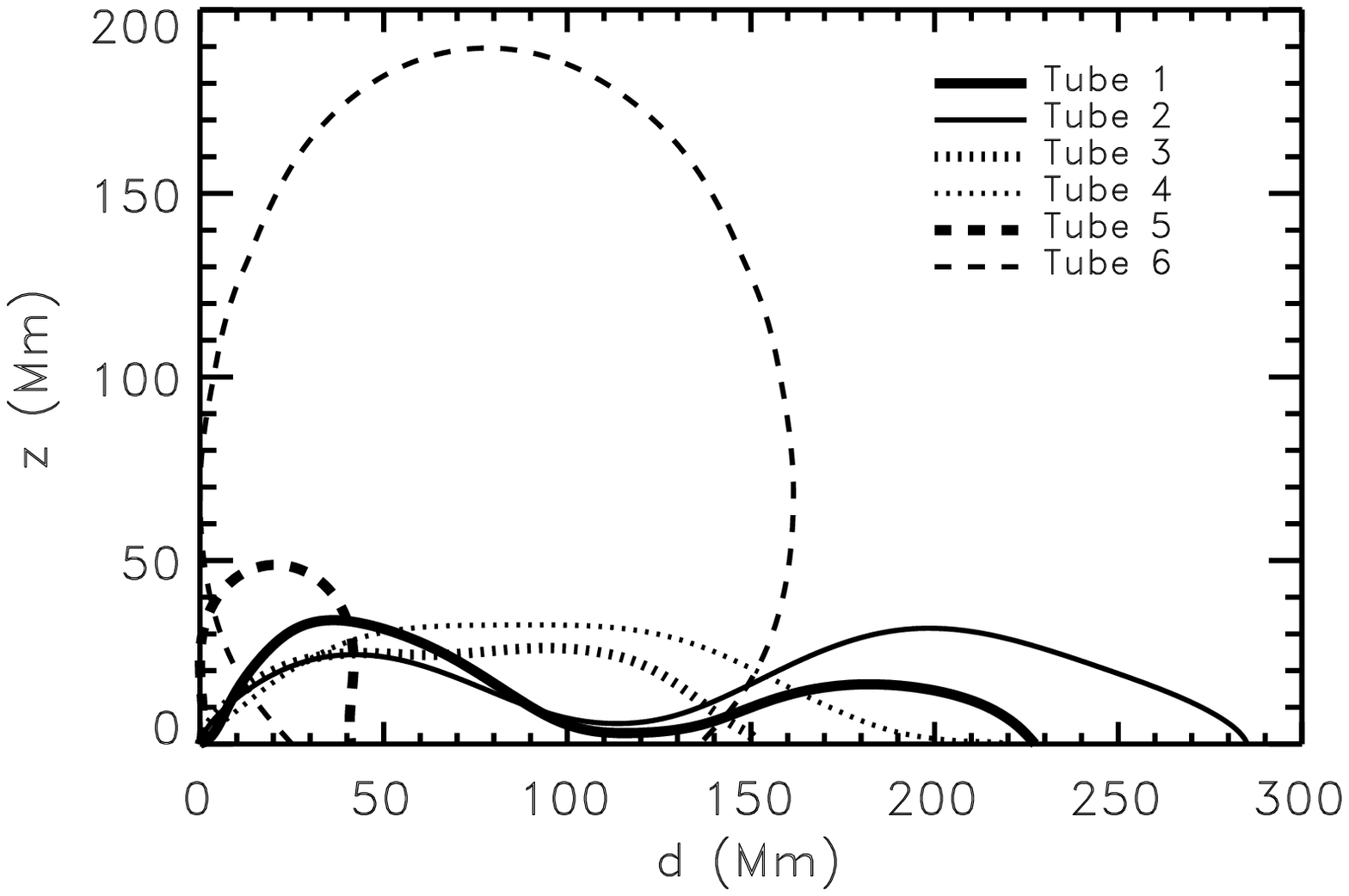}
\caption{Examples of the three types of flux tubes in our prominence model. The horizontal $d$ axis connects the two tube feet. Flux tubes with a deep dip (solid lines) are sheared lines, and form the filament channel core. Shallow-dip tubes (dotted lines) are also sheared lines, and cross the polarity inversion line above the deep-dip lines. Finally, the loop-like tubes (dot-dashed lines) have no dip, and form the overlying arcade structure.}
\label{repre-tubes}
\end{figure}

We performed the numerical simulations of the one-dimensional hydrodynamic equations with our code ARGOS (Adaptively Refined GOdunov Solver) with adaptive mesh refinement \citep[see][for details]{antiochos1999}. We have selected a discrete set of $125$ non-uniform flux tubes of the 3D sheared double-arcade magnetic structure shown in Figure \ref{magnetic_structure}. We have rescaled the computational domain of \citet{devore2005} to a box of 800 Mm parallel to the PIL, 200 Mm perpendicular to this PIL in the base plane, and 200 Mm in the vertical direction. The footpoints of the field lines were selected by imposing two uniform rectangular grids in the photosphere. The first grid is denser and covers the sheared region of the photosphere, being $40~\mathrm{Mm}$ wide centered at the PIL and $600~\mathrm{Mm}$ along the PIL. The second grid covers the outer region with no shear, and it is located only on one side of the PIL due to the symmetry of the magnetic field with respect to the PIL. This grid extends perpendicular to the PIL from the end of the first grid to $55~\mathrm{Mm}$ away. Along the PIL both grids have the same extension. The geometry of the 3D flux tubes is incorporated into our 1D simulations in two ways. The field line shape determines the projection of the gravity along the tube. In addition the non-uniform cross-sectional area $A$ is incorporated using conservation of magnetic flux along the tube. The position $s$ along the tube is described by the distance from one footpoint at the base of the corona. With this definition the positions of the two footpoints at the base of the corona are $s=0$ and $s=L$, where $L$ is the length of the coronal part of the tube. Setting the area at $s=0$ to be $A_{0}$, we can determine the area along the entire tube using $A(s) B(s) = A_{0} B_{0}$ where $B_{0}$ is the magnetic field strength of the sheared double arcade at $s=0$.

The geometry of the flux tubes in the whole magnetic structure can be classified roughly into three groups: tubes with deep dips, tubes with shallow dips, and tubes with no dips (loop-like tubes). In Figure \ref{repre-tubes} two examples of each group are plotted. The first group has deep dips; these are sheared, long, and low-lying lines with inverse polarity at their central parts (tubes 1 and 2 in Fig. \ref{repre-tubes}). The magnetic field and gravity jointly provide a stable equilibrium region in the dipped portion of the tube because the projected gravity forces the plasma to move to the center of the dip. The second kind of tube has a plateau shape with a central part almost horizontal, as we see in the examples 3 and 4 of Figure \ref{repre-tubes}, providing a region in the tube where the projected gravity is small or zero. These two groups of field lines belong to the sheared magnetic field, and they form the FC structure (Fig. \ref{magnetic_structure}). The third kind of tube is in the unsheared or weakly sheared arcade overlying the FC. These tubes are loop-shaped as exemplified by tubes 5 and 6 in Figure \ref{repre-tubes}. Most of the loops cross the PIL perpendicularly, but the short tubes near the FC structure are skewed from the perpendicular direction. Most loop-like tubes have a large projected gravity except in the tube apices. There is a small group of loop-like tubes at both ends of the FC with large inclinations (see Figs. \ref{magnetic_structure}a and \ref{magnetic_structure}b), with small projected gravity along the tube. Tube $5$ is a short loop that wraps around the inner core of the FC structure and is slightly skewed from perpendicular to the PIL. Tube $6$ is a very long loop reaching very high altitudes of around $190~\mathrm{Mm}$. Most of the selected $125$ tubes belong to the FC, and the others belong to the overlying arcade.

\begin{figure}[ht]
\centering
\includegraphics[width=12cm]{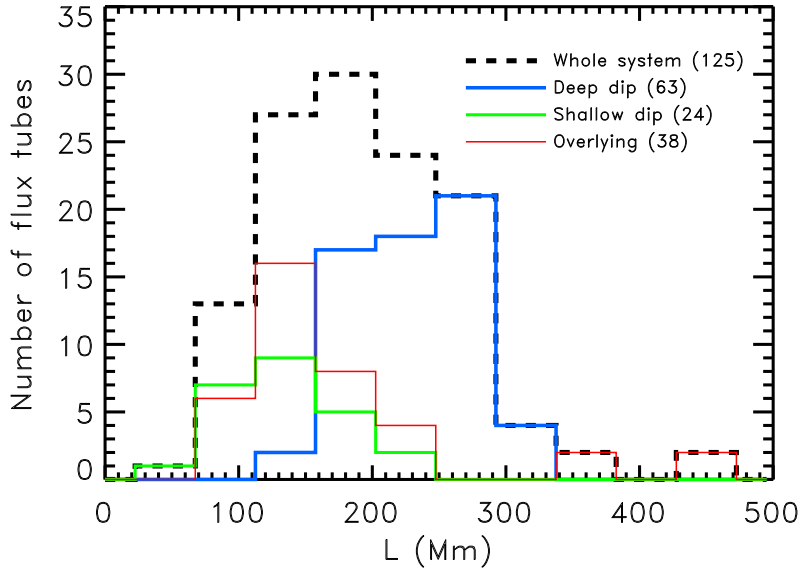}
\caption{ Histogram of the lengths of the selected flux tubes of the magnetic structure. The full set of $125$ (black dashed line) has a distribution similar to a Gaussian. The minimum length is $84~\mathrm{Mm}$ and the maximum is $479~\mathrm{Mm}$, with a mean value of $218~\mathrm{Mm}$. Tubes of the first kind (blue line) with deep dips have a Gaussian distribution with a mean length of $255~\mathrm{Mm}$. The second kind (green line), tubes with shallow dips, also have a Gaussian distribution but centered at $156~\mathrm{Mm}$. The third kind (red line), the overlying arcade loops, are tubes with a mean length of $192~\mathrm{Mm}$.}
\label{histogram}
\end{figure}

In Figure \ref{histogram} the histogram of the lengths of the flux tubes is plotted. The distribution has a Gaussian shape except for the contribution of very long tubes. The tube lengths are distributed between $84~\mathrm{Mm}$ and $479~\mathrm{Mm}$ with a mean value of $218~\mathrm{Mm}$. We compute also the contribution of the three kinds of flux tubes to the global distribution of tubes. We see that the right wing of the distribution (i.e., $L\ge218~\mathrm{Mm}$) is given mainly by the tubes of the first kind with deep dips; the main contribution to the left wing of the distribution comes from both shallow-dip and loop flux tubes. Thus, on average the tubes with deep dips are longer than the tubes with shallow dips. This is because lines in the first group are reconnected during the process of formation of the FC by interactions between the two sheared bipoles \citep{devore2005}. We have selected only a small number of tubes of the overlying arcade, as can be seen in the histogram. This is because we are interested mainly in the tubes of the FC core, but we also included a small sample of other tubes to take into account the evolution of the plasma outside the FC. The area expansion factors, $A(s)/A_{0}$, of the tubes of the first and second kinds in the corona are similar, with a mean value of approximately 3 in the central part of the tubes. In contrast, the area expansion of tubes of the overlying arcade has a mean value of 30 in the loop apices.

As in our previous studies, the volumetric heating has two components: a spatially uniform background heating, $H_{0}$, plus a localized component at each footpoint in the chromosphere, $H_{1}$ and $H_{2}$, that falls off exponentially in the corona. Thus,
\begin{equation}\label{heating-function}
H(s)=H_{0}+H_{1} ~e^{-s/\lambda}+H_{2}~e^{(s-L)/\lambda} ,
\end{equation}
where $\lambda$ is the heating deposition scale that we have set to $\lambda=10$ Mm, and the background heating is set to $H_{0}=1.5\times10^{-5}~\mathrm{erg}~\mathrm{cm}^{-3}~\mathrm{s}^{-1}$. The tube feet are heated asymmetrically, so $H_{1}$ and $H_{2}$ are different. We have set the larger value of the volumetric heating as $\mathrm{max}(H_{1},H_{2})=10^{-2}~\mathrm{erg}~\mathrm{cm}^{-3}~\mathrm{s}^{-1}$, and the heating asymmetry as $\mathrm{min}(H_{1},H_{2})/\mathrm{max}(H_{1},H_{2})=0.75$. In our $125$ independent simulations we have distributed randomly the position of the larger localized heating, so some tubes are more heated at $s=0$ and others at $s=L$. A $60\mathrm{-Mm}$ chromospheric region has been added to each flux tube footpoint in order to have a mass reservoir and sink for the coronal plasma evolution. In all flux tubes, an equilibrium situation was established first with only background heating $H_{0}$ turned on during the first $10^{5}~\mathrm{s}$. This small initial uniform heating populates the flux tubes with density, temperature, and pressure dictated by the scaling laws of \citet{Rosner1978}. After this $10^{5}~\mathrm{s}$, the localized heating was ramped up over $1000~\mathrm{s}$ and sustained at that level thereafter  for an additional $10^{5}~\mathrm{s}$. Further details of the numerical technique can be found in \citet{antiochos1999} and \citet{karpen2005}.

\section{Results}\label{results-sec}

The evolution of the plasma on each flux tube depends on the tube geometry and the asymmetry of the heating function \citep{karpen2003}. As we have seen in \S \ref{model-sec} the geometry of the field lines can be classified roughly into three groups. In Figure \ref{repre-tubes}, two representative tubes of each group have been plotted. In order to understand the process of formation of the whole prominence, we first study the evolution of the plasma on the selected set of six field lines.
\begin{figure}[!ht]
\centering
\vspace{-0.cm}
\includegraphics[width=9.cm]{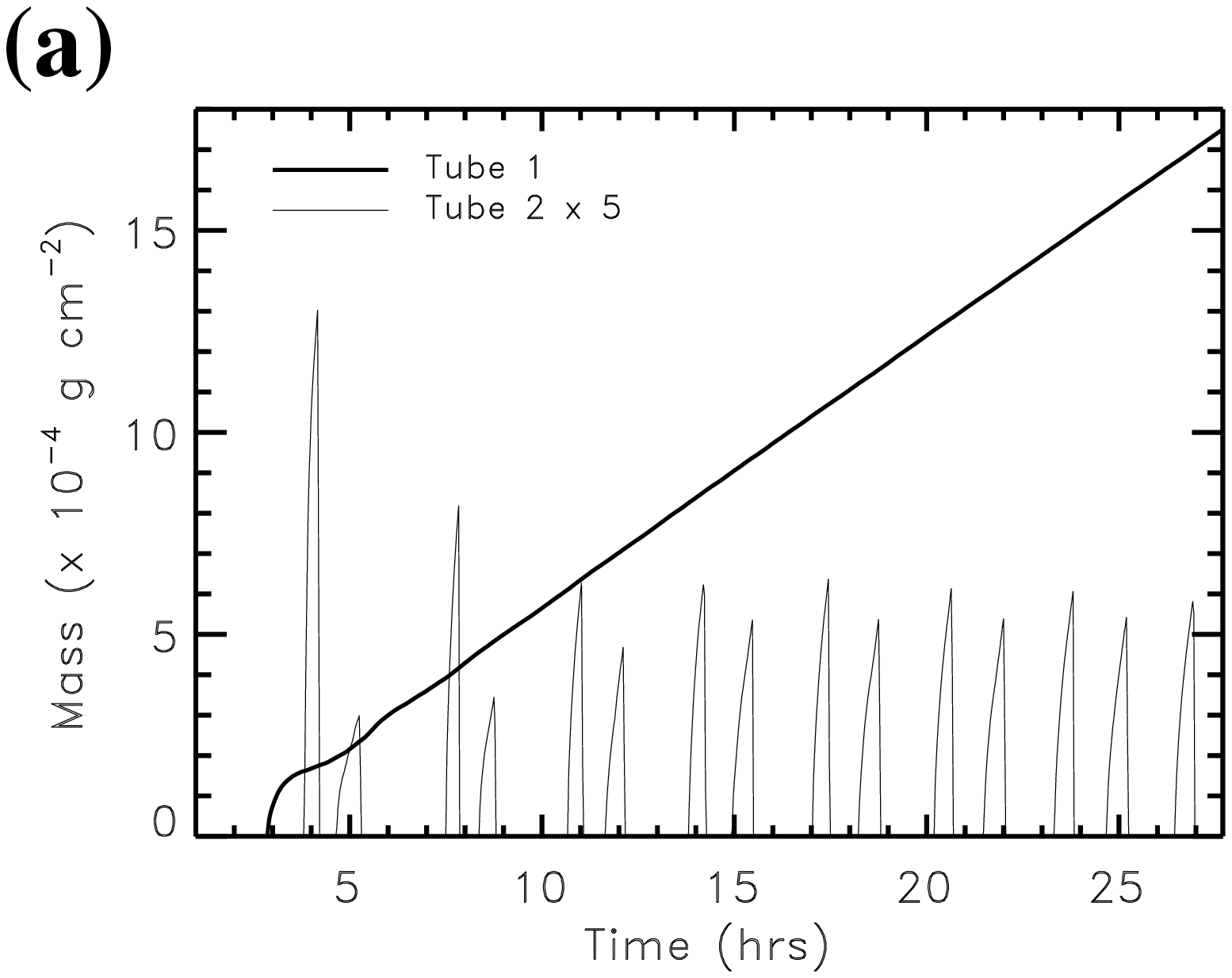}\vspace{-7.cm}
\includegraphics[width=9.cm]{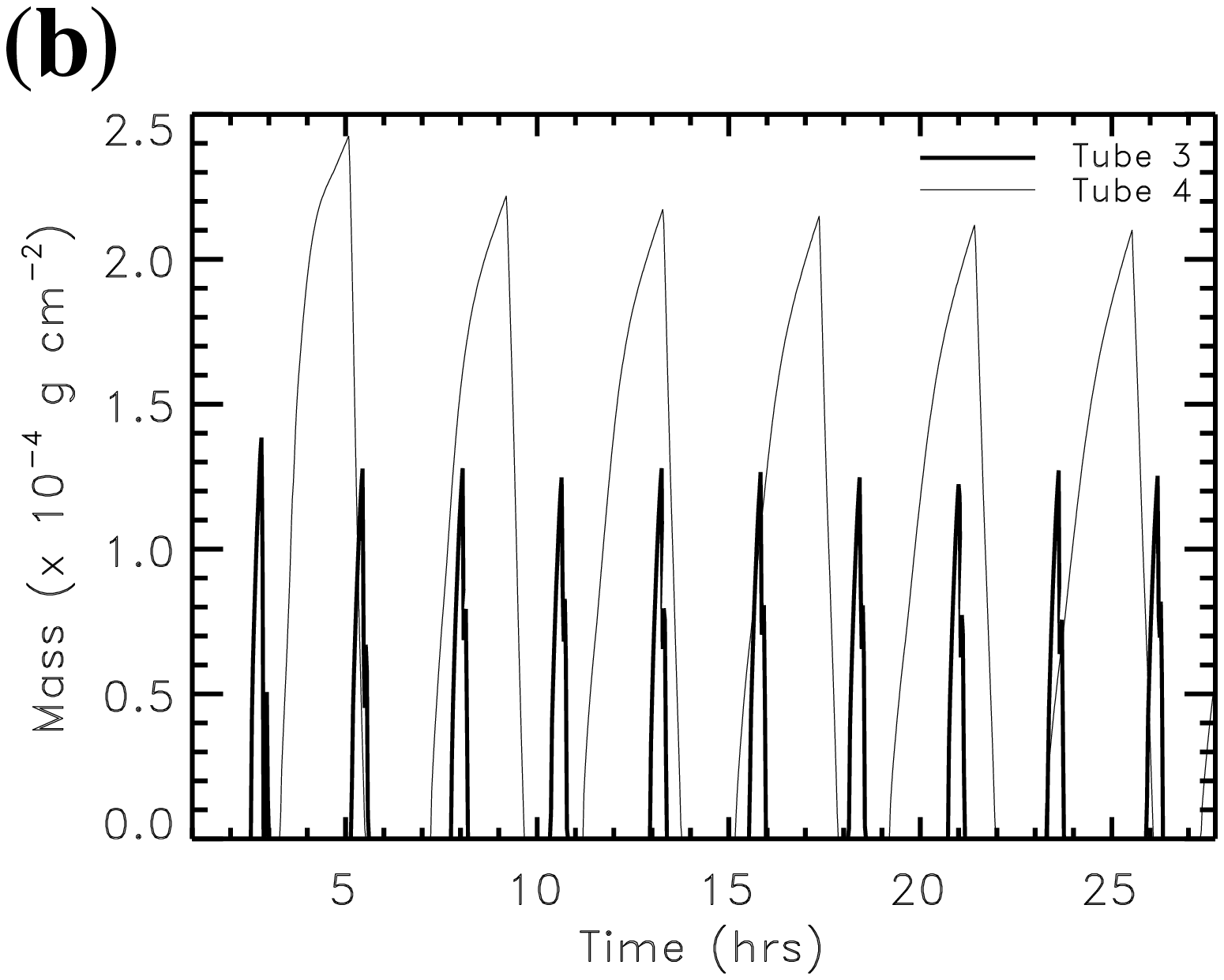}\vspace{-7.cm}
\includegraphics[width=9.cm]{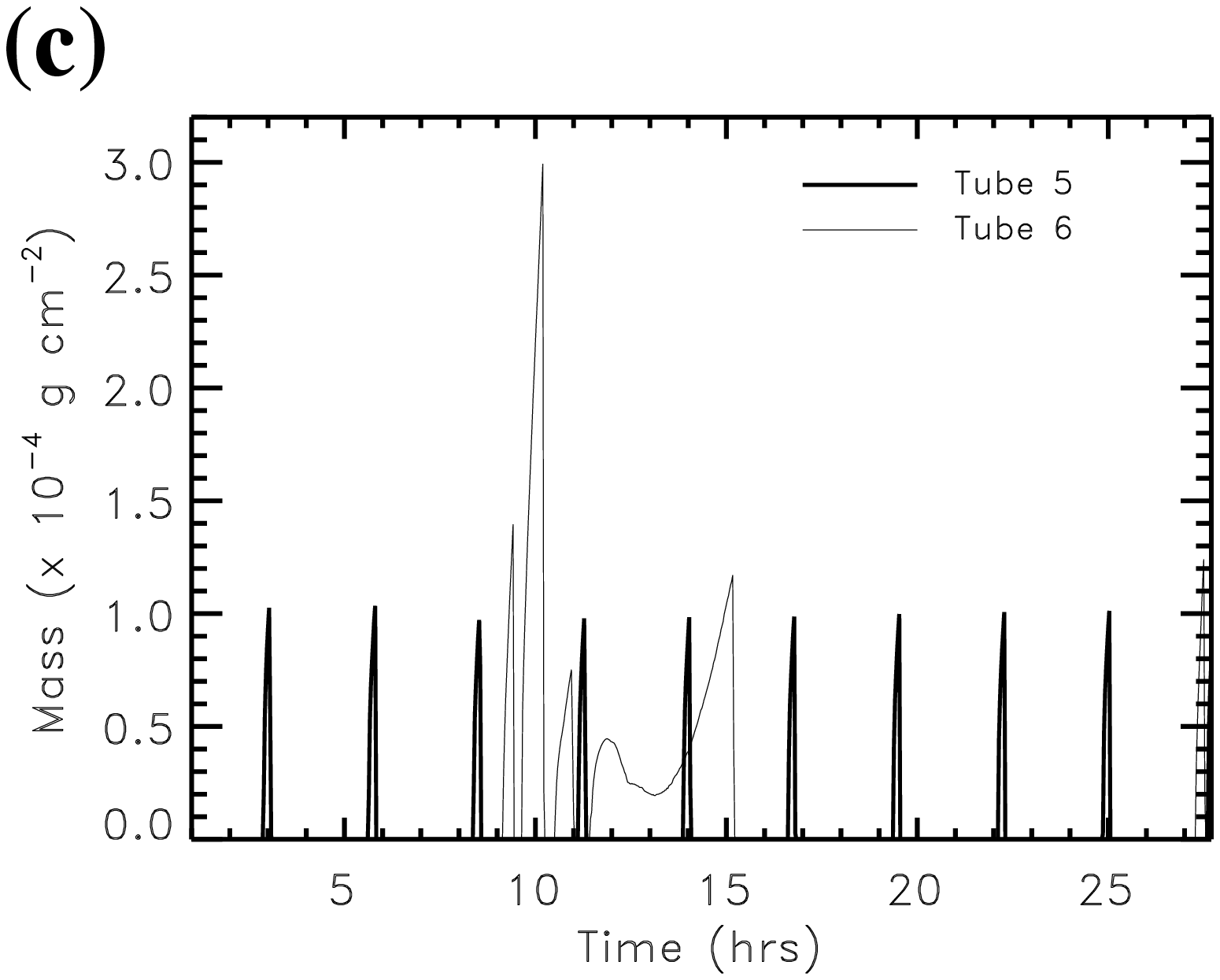}\vspace{-6.cm}
\caption{Temporal evolution of the condensation masses per unit area for the six representative tubes, where the masses are divided by $A_{0}$. {\bf (a)} Condensations in the two tubes with deep dips.  {\bf (b)} Condensations formed in tubes with shallow dips. {\bf (c)} Condensations in the two overlying arcade tubes.}
\label{individual-condensations}
\end{figure}
In Figure \ref{individual-condensations}, we have plotted the mass of the condensations as a function of time for the six simulation examples. Figure \ref{individual-condensations}a shows the evolution of cool condensations in tubes $1$ and $2$ with deep dips. The condensation in tube $1$ forms at $t=3~\mathrm{hrs}$ after the onset of the localized heating, after which the condensation mass grows linearly at a rate of $6.6\times 10^{-4} ~\mathrm{g}~\mathrm{cm}^{-2}~\mathrm{hr}^{-1}$ for the rest of the simulation. The condensation forms at the dipped portion of the field line and remains there while it accretes mass. Each condensation of tube $2$ shows completely different behavior, forming close to a footpoint and falling rapidly to it. This process is repeated in a limit cycle \citep{karpen2003,muller2003} with a period of $\sim3~\mathrm{hrs}$. In this cycle a pair of consecutive condensations forms, moves along the tube, and finally falls to the chromosphere. The difference between tubes $1$ and $2$ in terms of the plasma evolution is dictated by the heating asymmetry relative to the flux tube geometry \citep[see, e.g.,][]{karpen2003,karpen2005}. In both cases the heating is stronger in the righthand footpoints of Figure \ref{repre-tubes}. In tube $1$ the steep slope between $d= 40$ Mm and $110$ Mm keeps the condensation in the dip. However, in tube $2$ the slope is not steep enough to retain the condensation, so it falls to the lefthand chromosphere. 

The evolution of the mass in shallow-dip tubes, simulations $3$ and $4$, is shown in Figure \ref{individual-condensations}b. In these two simulation cases there are no stable condensations as in simulation $1$; only small and short-lived condensations are produced cyclically (note the change in scale between panels of Fig. \ref{individual-condensations}). In simulation $3$, the sequence of condensing and falling to the chromosphere is repeated every $2.5~\mathrm{hrs}$, and each condensation lives for approximately $0.5~\mathrm{hrs}$. However, simulation $4$ produces more massive condensations roughly every $4~\mathrm{hrs}$, each of which has a lifetime of $3~\mathrm{hrs}$. Tube $3$ has a length of $176~\mathrm{Mm}$, and tube $4$ has a length of $240~\mathrm{Mm}$. This indicates that the longer the loop, the larger the period of the limit cycle. 

Similarly, the evolution of the cool mass of the two loop-like tubes is plotted in Figure \ref{individual-condensations}c. Small condensations form in both loops of the overarching arcade. These two tubes ($5$ and $6$) have lengths of $121~\mathrm{Mm}$ and $479~\mathrm{Mm}$ respectively, with tube $6$ being the longest tube considered in this work. In simulation $5$, we see that each condensation forms quickly, lives for $\sim 0.25~\mathrm{hrs}$, then falls rapidly to the chromosphere. This process is repeated periodically every $3~\mathrm{hrs}$. The temporal evolution of the cool mass in loop $6$ is more complex, with several peaks reflecting a cluster of short-lived condensations that forms and disappears within an interval of $\sim 6$ hours. Only one entire condensation cycle occurs within the simulation time, but another condensation appears at the end of the run. Thus, the period of the evaporation/condensation cycle is around $18~\mathrm{hrs}$.

\subsection{Condensation masses and lengths}\label{mass-sec}

\begin{figure}[!ht]
\centering
\includegraphics[width=12.cm]{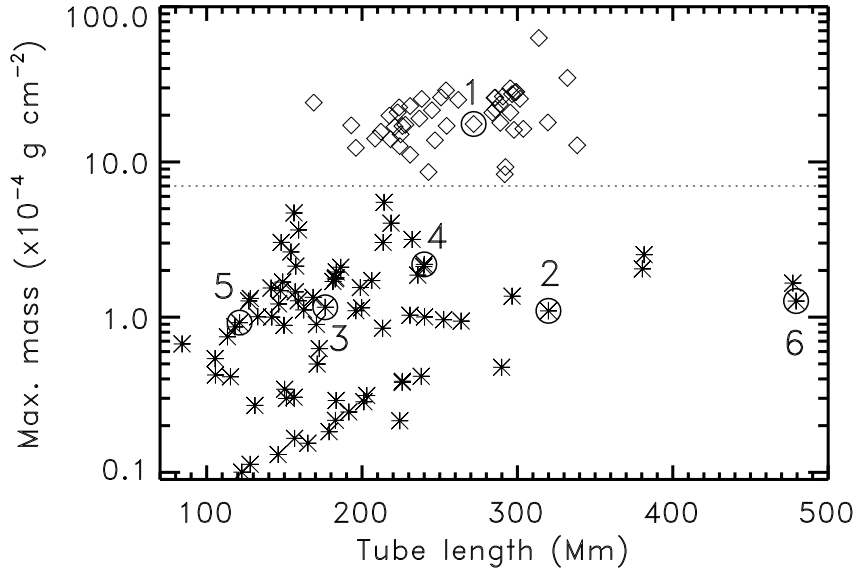}
\caption{Plot of the maximum mass of the condensation as a function of tube length. Diamond symbols correspond to condensations of the Population T (see text) with masses larger than the separator mass (dotted line). Asterisks correspond to condensations of Population B with masses smaller than the separator. The 6 representative tubes are numbered and surrounded with black circles (see text).
}
\label{max-mass-dip}
\end{figure}

Figure \ref{max-mass-dip} shows the maximum mass of the condensation in each flux tube. This quantity is useful for determining the ability of the tube to produce and accumulate cool mass according to its geometry. When the condensation mass grows monotonically, as in simulation $1$, this maximum mass coincides with the value at the end of the simulation. When $N$ consecutive evaporation/condensation cycles are produced, as in simulations $2$ to $6$, we average the maximum mass at each cycle as $M_\mathrm{max}=\sum_{1}^{N} m_{i}^\mathrm{max}/N$, where $m_{i}^\mathrm{max}$ is the maximum mass at $i$-th cycle. From this figure we see that the condensations can be divided mainly into two separate groups: condensations of relatively large masses (diamonds), which we denote Population T (for reasons to be explained shortly), and condensations of relatively small masses (asterisks), which we denote Population B. It is important to note that all $125$ simulated flux tubes studied here produce condensations. In the first group, condensations form and then grow monotonically until the end of the simulation, as for simulation $1$ (see Fig. \ref{repre-tubes}). In the second group, each condensation forms, moves along the tube, and finally falls to the chromosphere, as in tubes $2$ to $6$. We have found that approximately $7\times 10^{-4} ~\mathrm{g}~\mathrm{cm}^{-2}$ is the separator between the two condensation populations (dotted line). Condensations just above this limit initially have cycles of evaporation/condensation and fall to the chromosphere, but eventually the condensation mass grows monotonically. Condensations just below the limit start to grow monotonically but at some time the entire condensation falls to the chromosphere, and several cycles are produced in the simulation. All flux tubes that produce condensations of the first group are tubes with deep dips, but not all tubes with deep dips produce condensations of the first group, as discussed above for tube $2$. Small condensations of the second group are produced in all flux tube geometries, as the representative examples $2$ to $6$ indicate. There is no clear dependence of the condensation mass on the tube length. Condensations of the first group are produced in a range of tube lengths between $160$ and $340~\mathrm{Mm}$. This reflects the range of lengths of the deeply dipped field lines (see Fig. \ref{histogram}). In Figure \ref{max-mass-dip} we see a few examples of the second group of condensations for tube lengths larger than $350~\mathrm{Mm}$. These condensations are generated in the overlying arcade, and as mentioned in \S \ref{model-sec} we have only a small sample of those field lines. However, for our small sampling of the overlying arcade, the maximum condensation mass does not depend on tube length. In summary, all three kinds of tube geometries produce condensations of the Population B, with small masses, but a subset of deeply dipped flux tubes yields the massive condensations of the Population T.

\begin{figure}[!ht]
\centering
\includegraphics[width=12.cm]{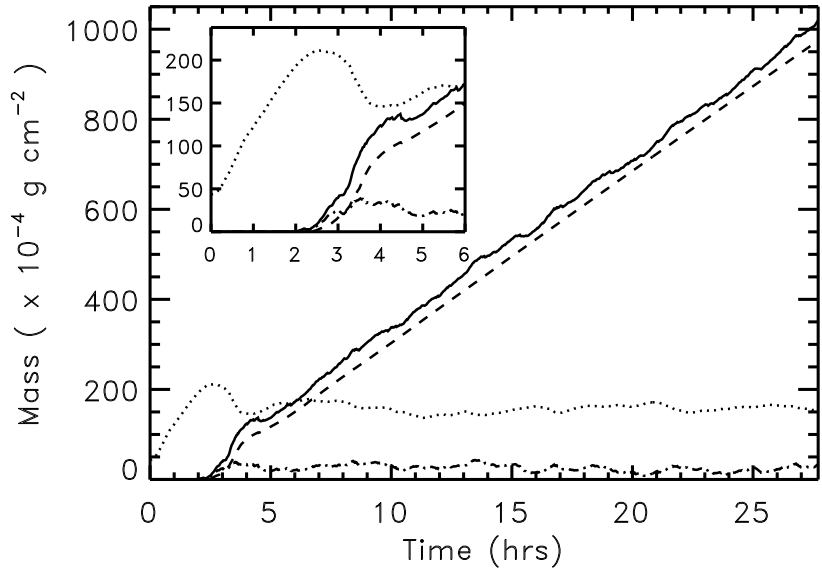}
\caption{Temporal evolution of the mass of the plasma in the filament channel. The mass of the condensations (solid line) is separated according to the contributions of Population T (dashed line) and Population B (dot-dashed line) condensations. The mass of the coronal plasma (dotted line) grows from an initial equilibrium value beginning at the localized heating onset ($t=0~\mathrm{hrs}$). The inset plot shows details of the early evolution.}
\label{totalmass}
\end{figure}

In Figure \ref{totalmass}, the mass of the hot plasma and the mass of the condensations in all tubes are plotted. The mass of the hot plasma (dotted line) grows steadily from a small value ($25\times 10^{-4} ~\mathrm{g}~\mathrm{cm}^{-2}$) given by the initial, uniformly heated, equilibrium. In this interval the initially underdense tubes are being filled with hot plasma as a consequence of the chromospheric evaporation at the tube feet. The growth of the hot plasma slows down at about $t=2.0~\mathrm{hrs}$, and finally stops at $t=2.5~\mathrm{hrs}$ (see inset plot of Fig. \ref{totalmass}).  In the interval $2.0$--$2.5~\mathrm{hrs}$, the chromospheric evaporation has increased the coronal density of most of the flux tubes and also their radiative losses, so the tubes start to cool down. At $t=2.0~\mathrm{hrs}$, the first condensations appear and the mass of the cool plasma starts to grow. It is interesting to note that in the interval $2.5$--$4.0~\mathrm{hrs}$ the coronal mass is reduced by $30\%$ of its maximum value. This partial evacuation of the corona occurs because the cool condensations accrete mass faster than the process of evaporation from the footpoints, due to a pressure deficit that pulls additional material into the condensations from the corona \citep[see][]{karpen2003}. The growth rate of the condensation mass is higher during this interval than in the rest of the simulation. After this time ($t=4.0~\mathrm{hrs}$) the hot mass reaches a more or less constant value of $158\times 10^{-4} ~\mathrm{g}~\mathrm{cm}^{-2}$, on average, for the rest of the simulation. Similarly, the condensation mass reaches a steady growth rate of $39\times 10^{-4} ~\mathrm{g}~\mathrm{cm}^{-2}~\mathrm{hr}^{-1}$. In this steady state, the rate at which mass condenses is balanced on average by the rate of evaporation from the footpoints of all tubes, so the mass of the hot coronal plasma remains more or less constant.

\begin{figure}[!ht]
\centering
\includegraphics[width=12.cm]{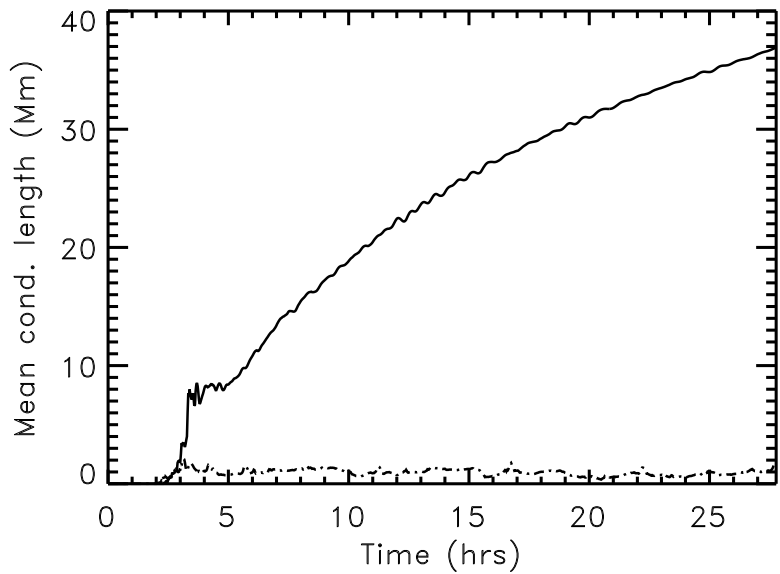}
\caption{The mean length of the Population T (solid line), and Population B (dot-dashed line) condensations as a function of time.}
\label{avelength-time}
\end{figure}

We have separated the contributions of the Populations T (dashed line) and B (dot-dashed line) to the total cool mass. Figure \ref{totalmass} shows that there is only a slight difference between the condensation mass of the entire system and the mass in tubes of Population T. This indicates that the main contribution to prominence mass comes from the Population T tubes. \citet{lin2005} estimated the observed width of the prominence threads to be $200~\mathrm{km}$. Assuming this value for the average diameter at the deep-dip tube centers, we can estimate the total mass of our model prominence to be $1\times10^{14}~\mathrm{g}$. This agrees with the observed values ranging from $1\times10^{14}~\mathrm{g}$ to $2\times10^{15}~\mathrm{g}$ \citep[see][for an extended review and references therein]{labrosse2010}. The mass estimate depends strongly on the volumetric filling factor of the cool plasma within the surrounding hot corona, which can be estimated by dividing the coronal volume occupied by cool threads by the total volume in which those threads are sparsely distributed. This total volume was determined as described in \S \ref{obser-sec} and is $\sim46\times10^{3}~\mathrm{Mm}^{3}$. The $47$ Population-T threads collectively occupy $\sim54~\mathrm{Mm}^{3}$, assuming cylindrical threads of $200~\mathrm{km}$ in diameter and the average thread length at the end of the run derived from Figure \ref{avelength-time}. Therefore the filling factor of our model is $\sim0.001$, which matches the lower limit of observed values \citep{labrosse2010}. To obtain a larger filling factor, as reported for some observations, either the threads are thicker than assumed (which would contradict the highest resolution data) or the prominence contains appreciably more threads than were selected for our modeling investigation.

\begin{figure}[!ht]
\centering
\includegraphics[width=12.cm]{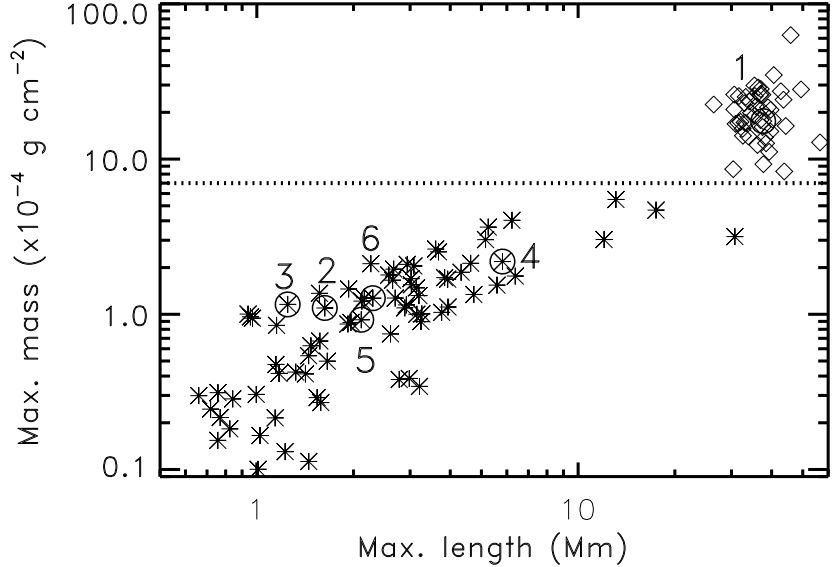}
\caption{The maximum mass of every individual condensation (see Fig. \ref{max-mass-dip} for symbol definitions) versus its length.}
\label{mass-length}
\end{figure}

Figure \ref{mass-length} shows the maximum condensation mass as a function of its maximum length. We see that condensations of the Population T are long, whereas condensations of the Population B are mainly short. This result is expected because, when a condensation grows by accreting mass, it also grows in size. For the Population T there is no clear relation between the mass and length of the condensations. Lengths are in the range $26.4$ to $56.6 ~\mathrm{Mm}$, with a mean value of $37 ~\mathrm{Mm}$. These large condensations form threads similar to those reported by  \citet{lin2005}; the computed lengths are in agreement with the observed lengths. We hereafter call the condensations of Population T ``threads''. Condensations of Population B are smaller, with a mean maximum length of $3.3 ~\mathrm{Mm}$. Hereafter, these small features are called ``blobs''. The advent of high-resolution high-cadence observations has revealed that such small, dynamic features exist throughout the corona \citep[e.g.,][]{schrijver2001,muller2003,degroof2004,muller2004,degroof2005}.

In Figure \ref{avelength-time}, the mean lengths of the two populations of condensations are plotted. The size of the threads (solid line) grows rapidly between $t=2.0$ and $t=3.5~\mathrm{hrs}$, when the mean size reaches $8~\mathrm{Mm}$. This period of time coincides with the partial evacuation of the coronal mass discussed in \S \ref{mass-sec}. Note also the sudden increase of the mean size between $t=3.0~\mathrm{hrs}$ and $t=3.3~\mathrm{hrs}$; in this $18\mathrm{-min}$ interval the mean length grows from $2$ to $8~\mathrm{Mm}$. Later on, the length remains more or less constant at about $8~\mathrm{Mm}$, until $t \approx 5~\mathrm{hrs}$ when the growth resumes up to the end of the simulation. In contrast to the total mass (Fig. \ref{totalmass}), the length does not grow linearly, and its growth rate decreases with time. Two factors contribute to the diminution of the growth rate, both of which are related to the geometry of the deep-dip flux tubes. First, after the formation of the thread the newly accreted plasma occupies regions of the tube with larger projected gravity than in the middle of the thread due to the concave-up shape of the dip. Then, accreted plasma significantly compresses the already condensed plasma because the pressure scale height in the condensation is small ($\sim 1~\mathrm{Mm}$). A similar result was reported by \citet{karpen2006}, who found that only nearly horizontal tubes can support extended condensations. Second, the newly accreted material fills regions of the tube that are wider than the central portion of the dip, due to the area expansion intrinsic to the sheared arcade mechanism for dip formation. In contrast to the threads, blobs rapidly reach a constant length of about $1~\mathrm{Mm}$. These small condensations can form anywhere in the tube and are short-lived (see Fig. \ref{individual-condensations}). This continual process of condensation and falling to the chromosphere keeps the temporally mean length more or less constant.

\subsection{Condensation velocities}\label{cond-vel-sec}

\begin{figure}[!ht]
\centering
\includegraphics[width=12.cm]{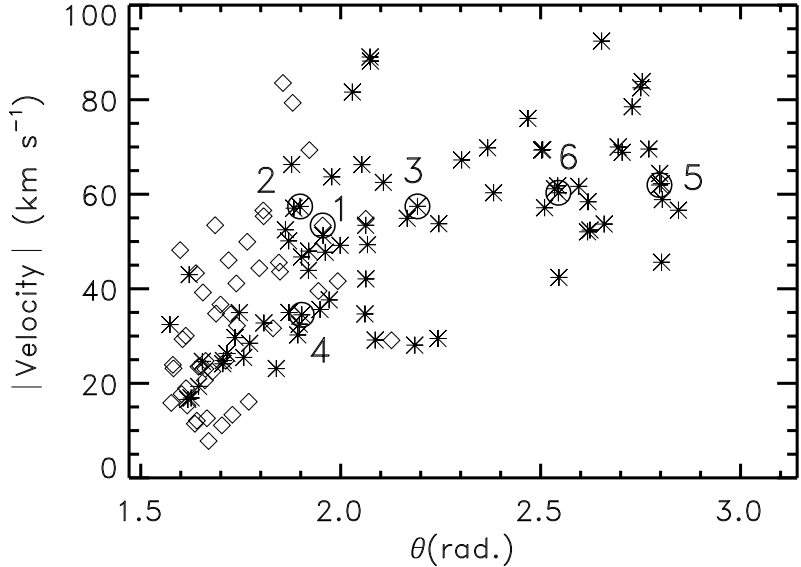}
\caption{Absolute values of the maximum bulk speeds of the condensations of both populations as a function of the angle between the bulk velocity vector and the vertical $z$-axis, $\theta$. The symbols are defined as in Fig. \ref{max-mass-dip}. The angle ranges from $\theta=\pi/2$ (horizontal motion) to $\theta=\pi$ (vertical downward motion).}
\label{velocities-fig}
\end{figure}
After formation a condensation moves along its field line in a manner determined by the local gravity along the tube and the localized-heating imbalance between the tube footpoints. Condensations are extended regions of cool and dense material in which each point moves with its own velocity. These motions can be decomposed into a bulk motion of the condensation plus local expansions or contractions. Here we investigate the dynamics of these individual condensations considering the bulk motion as the mass-averaged velocity in the condensation, i.e., the motion of the center of mass. In Figure \ref{velocities-fig}, the absolute value of the maximum bulk velocity of the condensations is plotted as a function of the angle between the bulk velocity vector and the vertical direction. Two kinds of motions are associated with the two condensation populations. Thread maximum velocities are concentrated in an angle range $\theta=\pi/2$--$2.0$ that is almost horizontal. Figure \ref{velocities-fig} reflects the fact that motion of the threads is in the concave part of the tube, and the motion in the dip is mainly horizontal with a small vertical projection, resembling a swinging motion. The direction of the velocities is not included in the figure, but in the movies associated with Figures \ref{front} and \ref{side} we see horizontal counterstreaming in adjacent threads. These condensations form in the dipped part of the tube, and fall to the bottom of the dip with a damped oscillatory motion \citep{antiochos2000,karpen2003}. The maximum velocity is reached beyond the bottom of the dip; this is why the angles are not exactly $\pi/2$ in Figure \ref{velocities-fig}. In contrast, blob velocities are in the range $\theta=\pi/2-2.9$, indicating that their motions cover the full span from nearly horizontal to nearly vertical. This result is expected because blobs are produced on flux tubes of all three kinds described in \S \ref{model-sec}, with geometries ranging from mostly horizontal tubes to mostly vertical tubes. The motion of the blobs on tubes with either deep dips or shallow dips (simulations 2 to 4) is relatively horizontal, while that of blobs formed in the overlying arcade (simulations 5 and 6) is nearly vertical. Most of the blobs slow down as they approach the tube feet, and in some cases several bounces occur before they fall finally to the chromosphere (see movies associated with the H$\alpha$ proxy of Figs. \ref{front} and \ref{side}). In general the maximum velocities are not reached at the tube feet. \citet{schrijver2001} reported similar motions in the coronal rain observed by TRACE, and suggested that the deceleration is the consequence of the pressure gradients of the underlying plasma close to the tube footpoints. \citet{muller2003} simulated this phenomenon and confirmed that blobs increase the pressure of the plasma below as they approach the chromosphere. Threads have a maximum velocity of $83.5~\mathrm{km~s^{-1}}$ and a minimum of $7.8~\mathrm{km~s^{-1}}$, with  a mean value of $34.2~\mathrm{km~s^{-1}}$. Blobs have a maximum velocity of $92.4~\mathrm{km~s^{-1}}$ and a minimum of $16.6~\mathrm{km~s^{-1}}$, with a mean of $51.0~\mathrm{km~s^{-1}}$. There is no clear dependence of the bulk velocity on the angle in the thread population. However, the blobs show a small tendency for the maximum bulk velocity to increase with the angle. This indicates that the downward velocities are larger in vertical tubes, but these velocities are far from the free-fall speeds. As we see in Figures \ref{front} and \ref{side}, the blobs form mainly below $30~\mathrm{Mm}$ in height. At this height the free-fall velocity is approximately $130~\mathrm{km~s^{-1}}$ \citep{Mackay2001,schrijver2001}, indicating that the condensations are slowed down by pressure forces impeding the free fall.

\section{Observational diagnostics}\label{obser-sec}

As in our earlier works on thermal nonequilibrium, we illustrate the dynamic behavior of the entire prominence model by computing synthetic images in selected representative EUV passbands. Unlike our prior studies of individual tubes seen from a single point of view, these images reflect the 3D structure of the prominence model with orientation and line-of-sight integration effects included. We have assumed that the emission in the EUV channels is optically thin and computed the images using the temperature response function of the Atmospheric Imaging Assembly (AIA) instrument aboard the Solar Dynamics Observatory (SDO) satellite. We show the AIA band passes 171\AA, and 211\AA. Several spectral lines contribute to each EUV passband, associated with different ions and ionization degrees. The peak temperature of the 171\AA\ passband is $\log\,T=5.8$, and the main contribution comes from Fe IX. The 211\AA\ band has a peak temperature of $\log\,T=6.3$, and the main contribution is from Fe XIV. In addition, a proxy for the H$\alpha$ emission is computed as in our earlier papers \citep[see, e.g.,][]{karpen2001}, setting the intensity constant whenever $T\le35,000~\mathrm{K}$ and zero at higher temperatures. We assume that the flux tubes are unresolved by the telescope, and use the AIA instrument's point spread function half-width of 0.6''.

\begin{figure}[p]
\centering
\includegraphics[width=8cm]{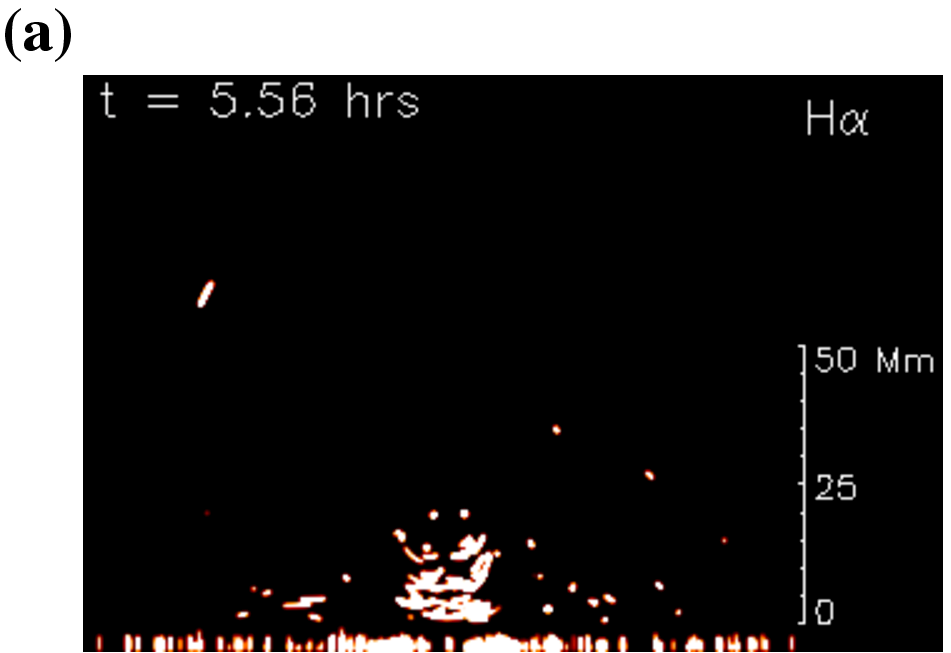}\includegraphics[width=8cm]{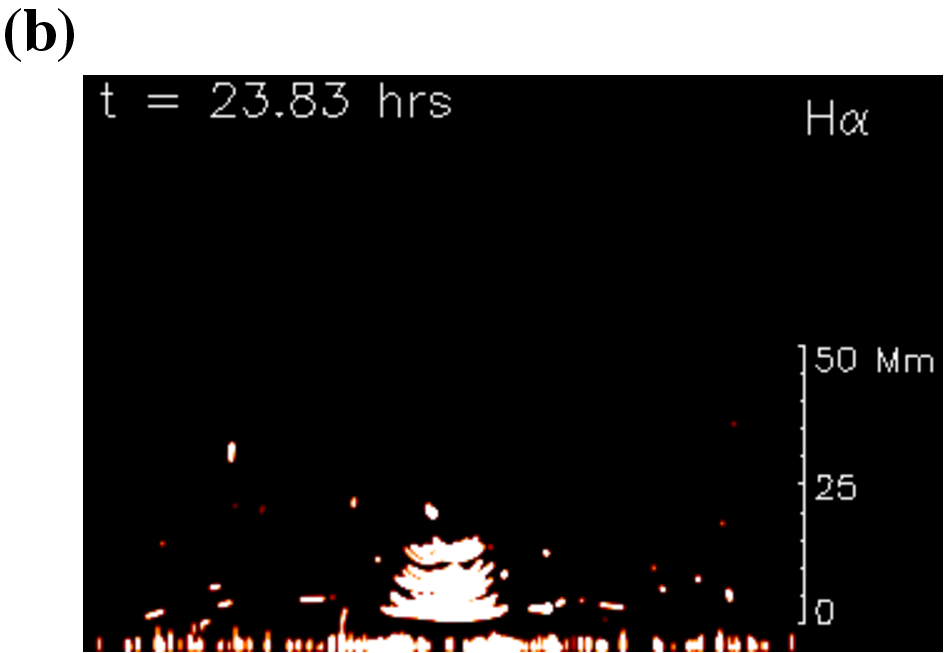}
\includegraphics[width=8cm]{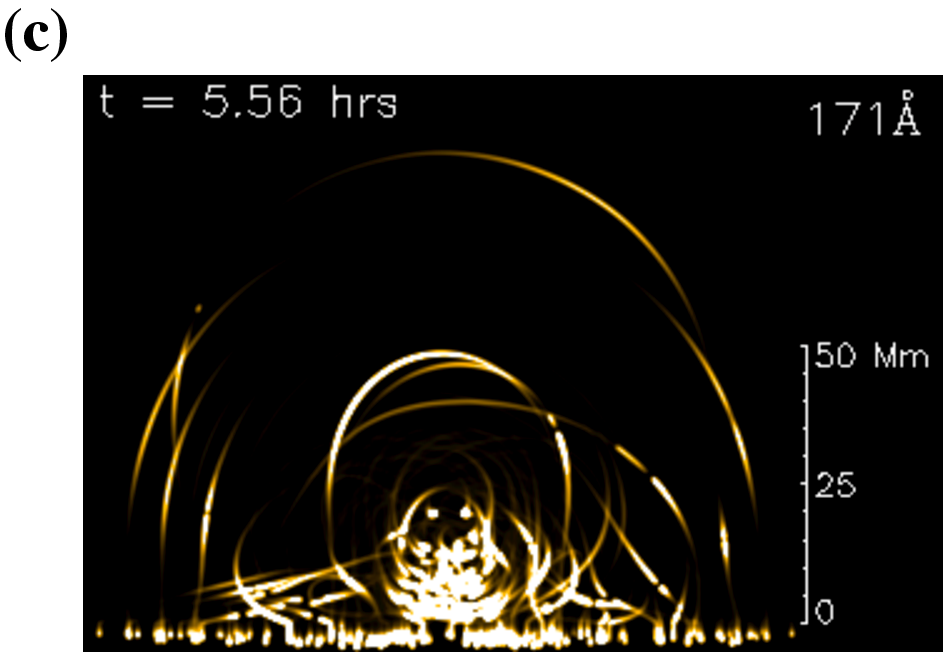}\includegraphics[width=8cm]{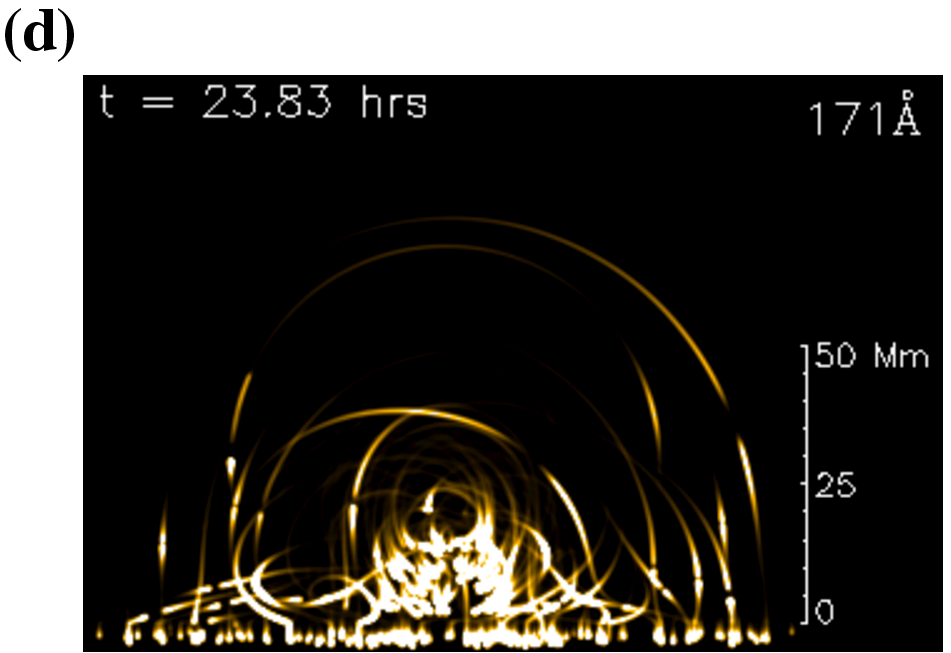}
\includegraphics[width=8cm]{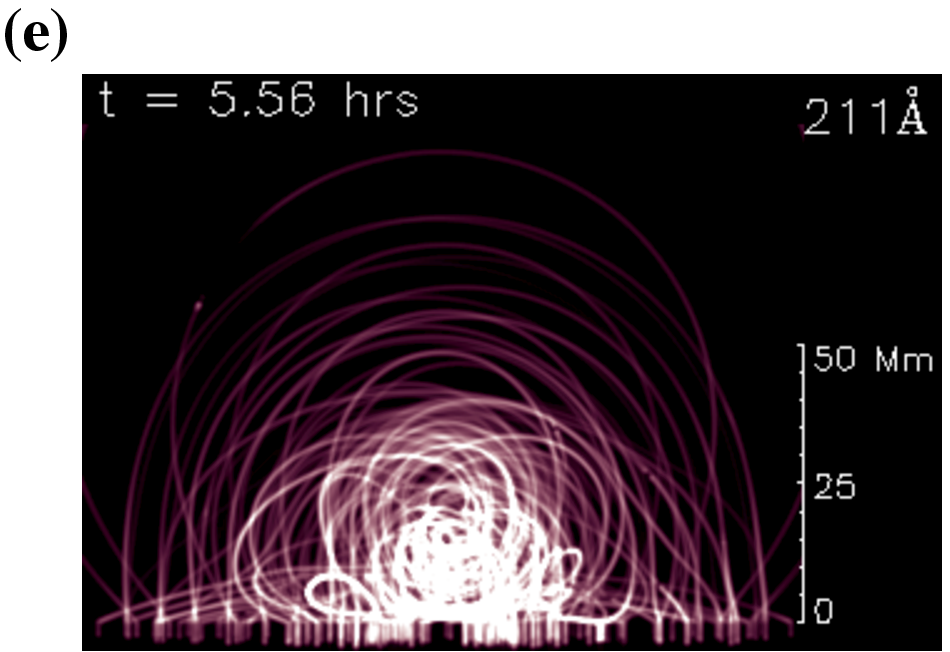}\includegraphics[width=8cm]{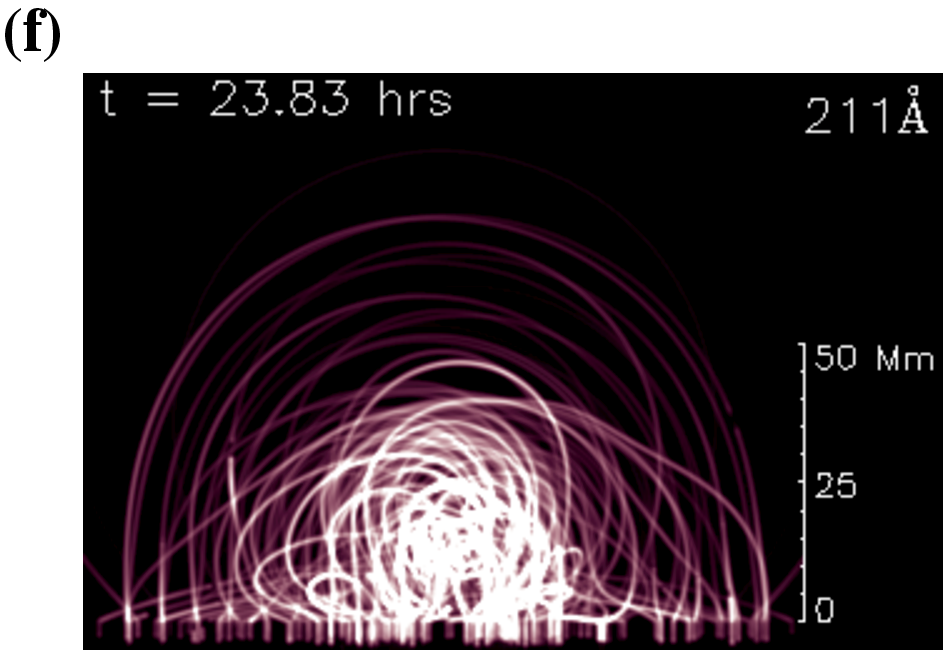}
\caption{Time evolution of the whole system in end-view with different filters. In {\bf (a)} and {\bf (b)} the H$\alpha$ proxy is shown at $t=5.56~\mathrm{hrs}$ and $t=23.81~\mathrm{hrs}$, respectively. Similarly, {\bf (c)} and {\bf (d)} show these two stages of the evolution in the 171\AA\ passband of the AIA/SDO instrument. In {\bf (e)} and {\bf (f)} the temporal evolution in the 211\AA\ emission intensity is plotted. (An animation of this figure is available in the online journal.)}
\label{front}
\end{figure}

\begin{figure}[p]
\centering
\includegraphics[width=15.cm]{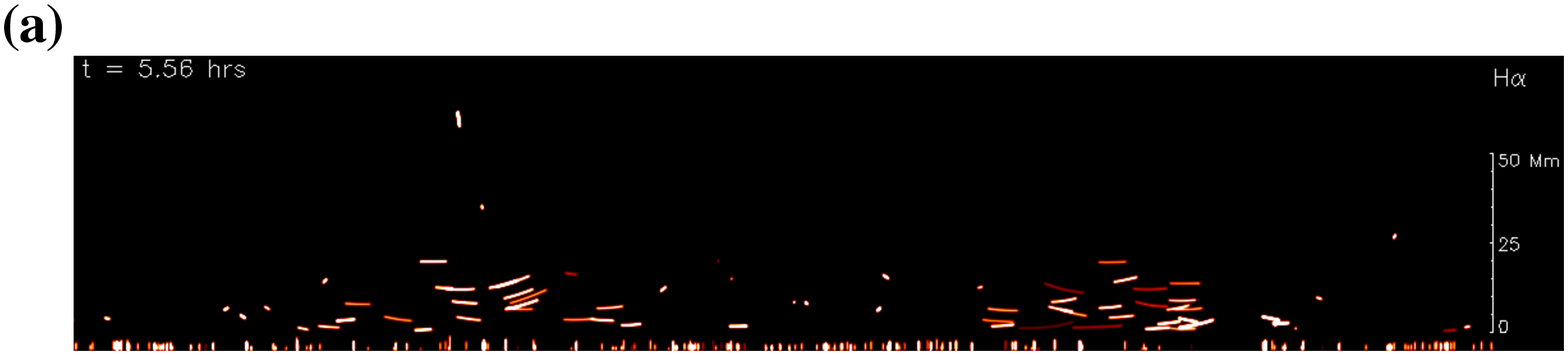}
\includegraphics[width=15.cm]{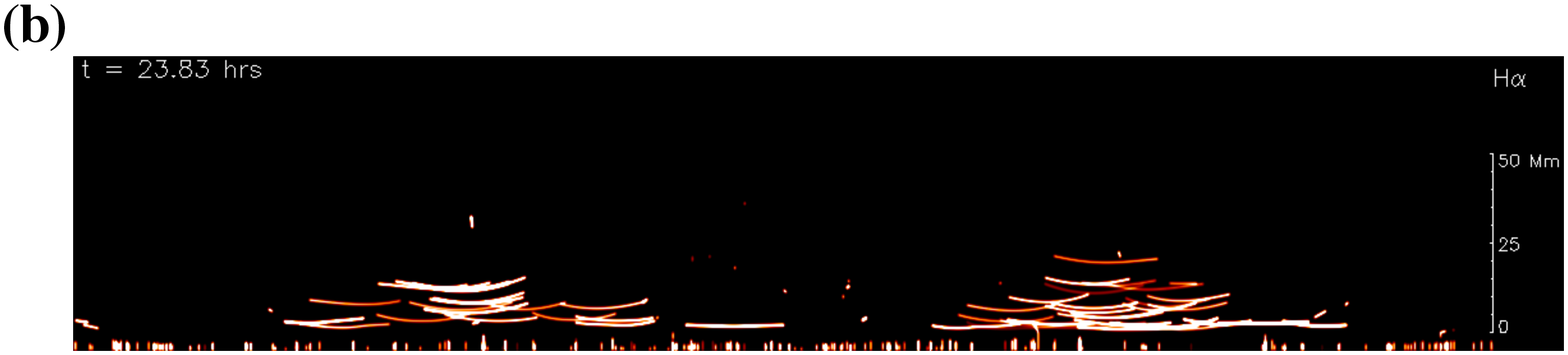}
\includegraphics[width=15.cm]{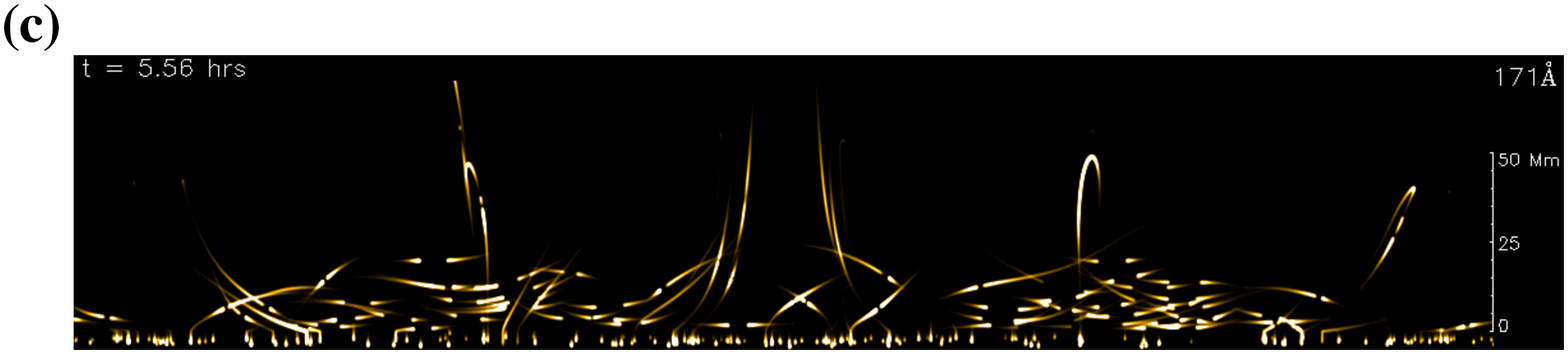}
\includegraphics[width=15.cm]{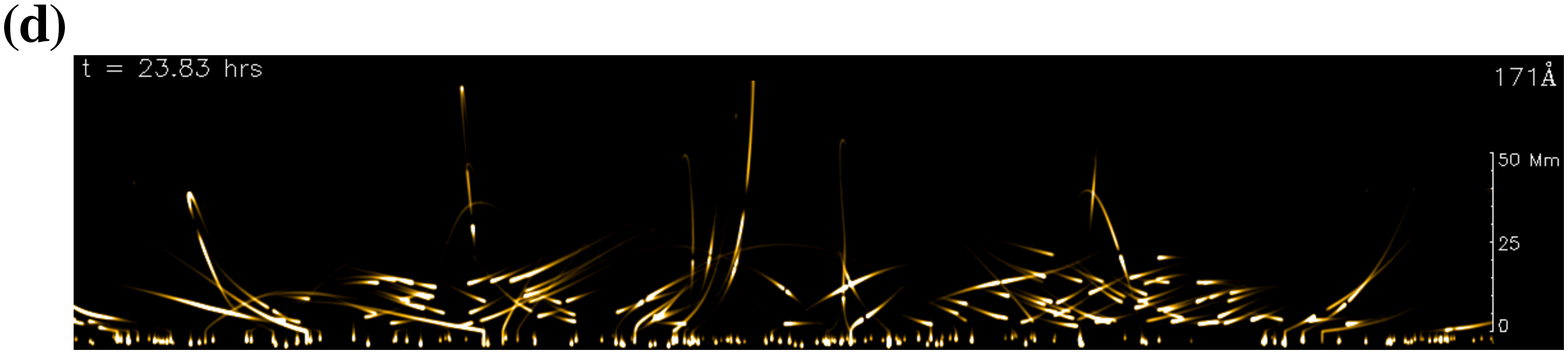}
\includegraphics[width=15.cm]{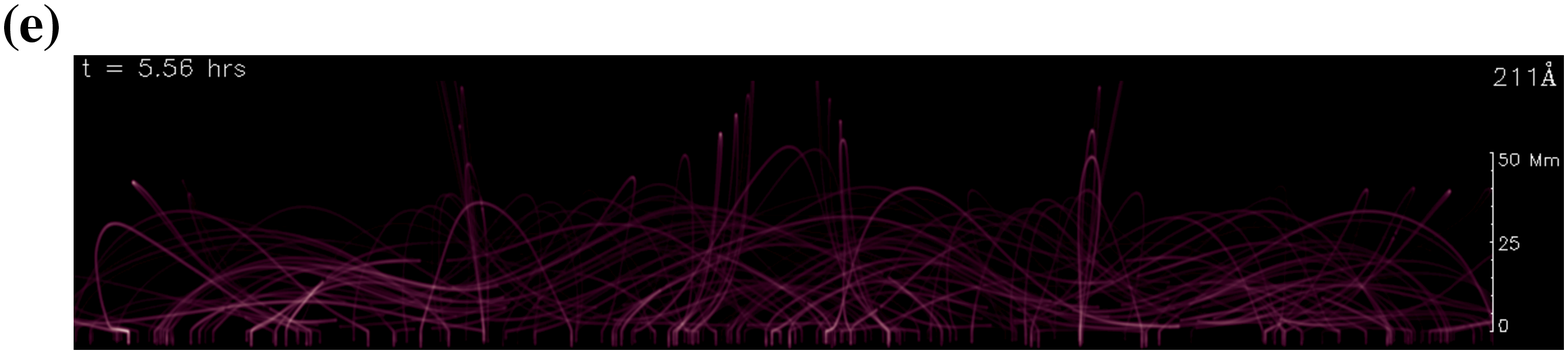}
\includegraphics[width=15.cm]{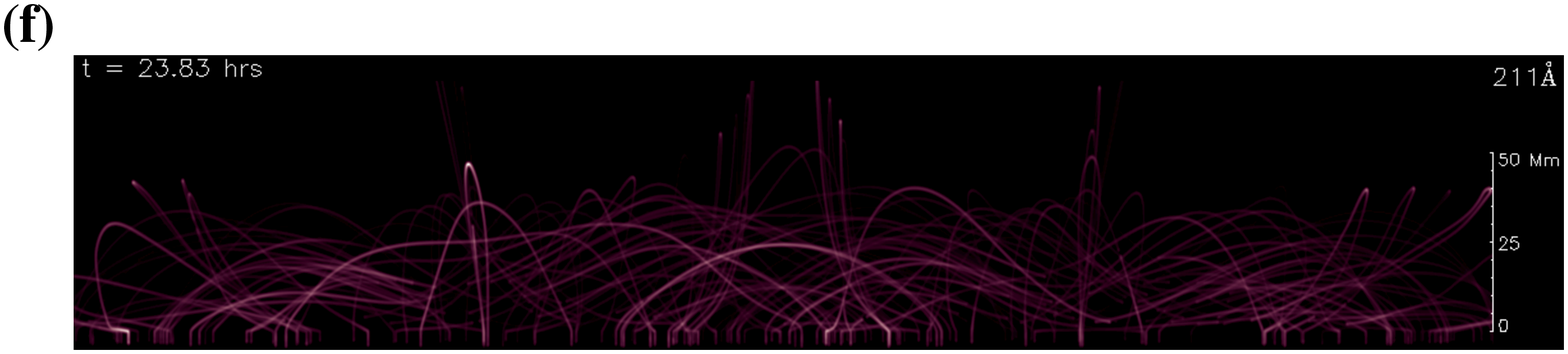}
\caption{Same as Fig. \ref{front} showing the temporal evolution in emission intensities in the three filters, but from the side-view orientation. (An animation of this figure is available in the online journal.)}
\label{side}
\end{figure}

We have constructed off-limb images in two perpendicular orientations of the system relative to the line of sight (LOS). In the end-view orientation the LOS is parallel to the PIL or filament-channel axis. In the side view the relative orientation of the LOS is perpendicular to the PIL. Figures \ref{front} and \ref{side} show the temporal evolution in all of the considered channels in the end and side views, respectively. In both views different types of structures are visible in different emission lines. In the H$\alpha$ channel we see the temporal evolution of the cool condensations, and distinguish between the two condensation populations. As we described in \S \ref{mass-sec} the threads are the largest cool condensations, containing most of the prominence mass. From the H$\alpha$ images we identify two separated prominence segments in the sheared double arcade. Both have a trapezoidal shape with a base of approximately $20~\mathrm{Mm}$ width (end view), $115~\mathrm{Mm}$ length (side view), and $20~\mathrm{Mm}$ height. Threads form in an off-center position in the dipped portion of the flux tubes, and oscillate around the bottom of the dips. In Figure \ref{side}a the recently formed threads are shown. The mean length of the threads is approximately $10~\mathrm{Mm}$ (see Fig. \ref{avelength-time}), and they move mainly horizontally with a large amplitude. In the movies accompanying the figures we clearly see the opposite motions on adjacent threads. \citet{antiochos1999} proposed this mechanism of condensation and oscillation in the dips of the flux tubes to explain the counterstreaming motions reported by \citet{zirker1998}. Here we clearly see how a bundle of threads manifests counterstreaming motions. These motions are also present in the end-view orientation (Fig. \ref{front}b). However, the direction of the motion is slightly different with respect to the LOS and the projected amplitude of motion is small. In Figure \ref{side}b the threads are larger than in the previous frame \ref{side}a, with a mean length of $35~\mathrm{Mm}$. We clearly see the concave-up shape of the threads in H$\alpha$ (Figures  \ref{front}b and \ref{side}b) lying in the bottom of the dips. The amplitude of the oscillations is now much smaller than in the initial stages of the oscillation, because of the strong damping reported by \citet{antiochos2000}.

Condensations of the population B (blobs) are small features seen in the H$\alpha$ synthetic images in all regions of the filament channel and the overlying arcade. These ubiquitous features form mainly at heights below $50~\mathrm{Mm}$, although some blobs form at greater heights up to $80~\mathrm{Mm}$.  Many blobs condense and move along the field lines between the two thread pyramids, producing a counterstreaming flow of blobs in that region (see the accompanying movie of Figs. \ref{side}a and \ref{side}b). Blobs forming in the overlying arcade rapidly fall to the chromosphere with an accelerated motion following the magnetic field. As we pointed out in \S \ref{cond-vel-sec}, these motions are not free-fall movements.  When blobs approach the tube footpoint, their movement decelerates before they enter into the chromosphere.

Figures \ref{front}c and \ref{front}d show synthetic images in the 171\AA\ passband in the end-view orientation. In this channel the emission comes from relatively warm plasma at $\log\,T=5.8$. In these figures, we clearly see strong emission from the core of the filament channel, where the prominence resides. We say that the emission from a region or a feature is bright or strong when the intensity is larger than three times the standard deviation of the intensities of all pixels of the entire image. The source of this emission is the warm plasma in the prominence-corona transition region (PCTR) at both ends of the individual threads and blobs (see the corresponding H$\alpha$ images). In this orientation the maximum intensity per image averaged over the simulation time is $637~\mathrm{DN\, pix^{-1}\,s^{-1}}$. The emission in this wavelength is strong in the core because the threads are concentrated in this region and the bright features are more or less aligned with the LOS. 

In the side views (Figs. \ref{side}a and \ref{side}b) the emission from the filament core is not as strong, with a temporal average of $374~\mathrm{DN\, pix^{-1}\,s^{-1}}$, because the bright features are less concentrated and not aligned with the LOS. However, from this point of view we clearly see the bright structures resembling two opposing tadpoles (or comets with tails) on both ends of the threads. Between each pair of tadpoles is an emission gap associated with the cool material of the condensation. These tadpole features are the consequence of the temperature and density distributions at the ends of the condensations. The temperature rapidly rises from several thousand K near the condensation to several million K in the hot parts of the tube, while the density inversely declines. In an intermediate region between the condensation and the corona, the temperature and density conditions are appropriate to produce the bright emission. The gradients of the temperature and density steepen close to the condensation and flatten farther away, explaining the tadpole shape of the emission. These bright features move together with the condensations; therefore, as shown in the accompanying movies, counterstreaming motions are also evident in the 171\AA\ passband. 

The end-view images reveal that the emission from the outer region of the prominence is fainter than from the filament core. In this region, the bright features are associated with Population B blobs. Since blobs are small condensations, the emission gap between their pairs of tadpoles is smaller than in case of the threads. The emission depends on the motion of the blobs: the leading tadpole is brighter than the trailing tadpole, relative to the direction of the blob motion, and in some cases the trailing tadpole is very faint. The animations also reveal long, transient brightenings, which in some cases illuminate most of the tube. These very long, bright features are consequences of the evaporation/condensation cycles. Before a condensation is produced, a large portion of the material of the flux tube cools down as the density increases. When the temperature of this denser plasma is approximately the peak temperature of the filter, a strong brightening is produced in that passband. The plasma continues to cool down, and finally the condensation is produced. At this moment the long, bright feature disappears and the tadpoles appear in 171\AA. All of the flux tubes in the filament channel produce such long features. However, the tubes with the Population-T condensations produce this transient brightening only in the initial stages of the simulation, just before thread formation (around $t=2~\mathrm{hrs}$). Thus, we expect large-scale transient brightenings to occur in the 171\AA\ passband, both in the core of the FC and in the overlying arcade.

Figures \ref{front}e and \ref{front}f illustrate the temporal evolution of the system in the end-view orientation for the hottest channel, 211\AA. The temporally averaged maximum emission in this passband and orientation is $52~\mathrm{DN\, pix^{-1}\,s^{-1}}$. In these images we see that hot plasma fills most of the FC structure, without the small features that dominate the 171\AA\ images, because the 211\AA\ channel has a wide temperature response function \citep[see, e.g.,][]{lemen2011}. The hot material is evaporated from the chromospheric plasma by the footpoint heating. In both images, the core of the structure is bright, resembling the ``chewy nougats'' observed by Yohkoh \citep{hudson1999,hudson2000}. In contrast with the 171\AA\ images, however a diffuse halo surrounds the prominence in 211\AA\ emission reaching a height of $40~\mathrm{Mm}$, resembling the ``hot  prominence shrouds'' reported by \citet{habbal2010}. In the side views (Figs. \ref{side}e and \ref{side}f) the emission is more uniform, with no distinguishable bright structure in the core. In this orientation the time-averaged maximum emission is $28~\mathrm{DN\, pix^{-1}\,s^{-1}}$. This demonstrates that the intensity distribution depends on the relative orientation of the FC to the LOS. In particular, the bright core appears when only the PIL is nearly aligned with the LOS. In the end view, the flux tubes in the core are closely aligned with the LOS, producing the strong core emission. In the side views (Figs. \ref{side}e and \ref{side}f), the 211\AA\ emission is more uniform and the core is darker because the cool plasma of the prominence emits little in this bandpass, an effect also known as emissivity blocking \citep{anzer2007}. In addition, the emission in this channel is more or less constant throughout the simulation time (see the accompanying movies) in both orientations. This indicates less dynamism of the hotter plasma than the cooler plasma. 

\subsection{Differential emission measure}\label{dem-sec}

\begin{figure}[!ht]
\centering
\includegraphics[width=12.cm]{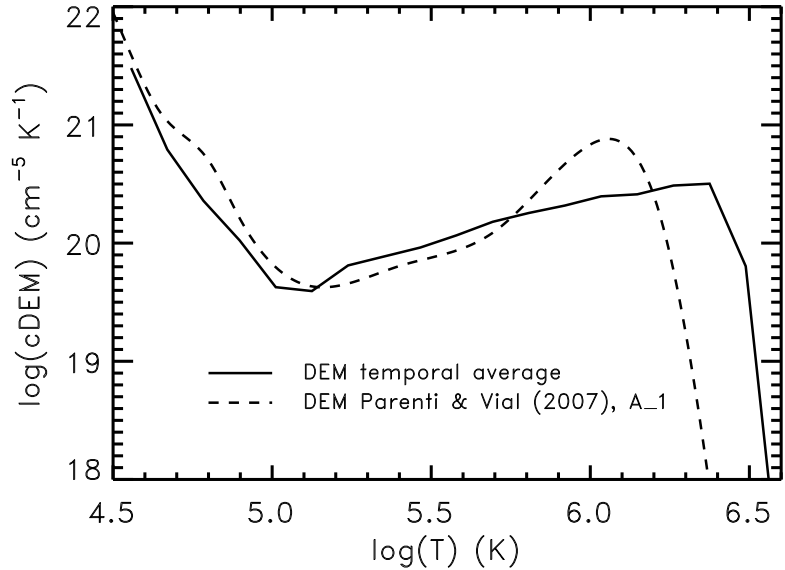}
\caption{Time-averaged differential emission measure vs. temperature for the whole system of flux tubes of the filament channel and the corresponding overlying arcade (solid line). The observed best fit obtained by \citet{parenti2007} for prominence section A\_1 (data courtesy of Parenti 2007 [private communication]) also is shown (dashed line).}
\label{dem}
\end{figure}

In Figure \ref{dem}, the differential emission measure (DEM) of our prominence model and overlying arcade is shown, taking into account all of the flux tubes of the filament channel structure. This is averaged in space and time for each flux tube, and then combined over the whole volume. To avoid the contribution of the initial transient phase, we averaged the DEM from $5~\mathrm{hrs}$ after the onset of the localized heating to the end of the simulation. The DEM is computed as in \citet{karpen2008}. The result of this computation is normalized by a prescribed column depth, which is chosen to ensure a good fit with the DEM values derived from prominence observations by SOHO SUMER \citep{parenti2007}. The resulting column depth is $h \approx 10~\mathrm{km}$, two orders of magnitude below our previous results ($h\approx1000~\mathrm{km}$) because the present work includes approximately 100 times as many tubes. As we see in the figure, our model reproduces very well the shape of the observed quiet-prominence DEM between $\log\,T = 4.6$ and $5.7$. In addition, the location of the relative minimum of our profile is $\log\,T = 5.1$, consistent with the observational value. As argued in \citet{karpen2008}, the additional emission at higher temperatures ($\log\,T \approx 6$) can be attributed to the unsubtracted background and foreground coronal contribution to the observed DEM of \citet{parenti2007}. Moreover, in our set of selected flux tubes we have considered only a small fraction of the overlying arcade loops that exist in a realistic structure. These loops of the overlying arcade contribute mainly to the warm and hot part of the DEM function because the condensations formed in those loops are small. Therefore, increasing the  number of overlying arcade loops should yield better agreement between the computed DEM and the observed curve at temperatures higher than $\log\,T = 5.7$. We plan to extend our model to include more of the overlying arcade.

\section{Discussion and conclusions}\label{discussion-sec}

We have developed a comprehensive three-dimensional model of the plasma dynamics and energetics of a prominence and its overlying arcade. We have found that condensations are formed in all of the flux tubes studied. There are two populations of condensations characterized by their cool mass. Repeated cycles of condensations forming and falling to the chromosphere produce condensations of masses below $7\times 10^{-4} ~\mathrm{g}~\mathrm{cm}^{-2}$, and small sizes. Condensations that linger in the dipped portion of the tube have a constant growth rate, and reach large masses above this limit and large lengths. We have called the small condensations ``blobs'' and the large ones ``threads''. Blobs have a mean length of about $1~\mathrm{Mm}$, while threads are larger condensations with a mean length of $37~\mathrm{Mm}$. The dominant contribution to the prominence mass comes from the threads. This mass grows at a constant rate of $39\times 10^{-4} ~\mathrm{g}~\mathrm{cm}^{-2}~\mathrm{hr}^{-1}$. Assuming that the threads have a width of $200~\mathrm{km}$, the total mass of our model prominence is $1\times10^{14}~\mathrm{g}$ at the end of the simulation, in agreement with observed values for small prominences. When threads are formed, they oscillate around an equilibrium position; this motion is mainly horizontal (parallel to the solar surface), with a mean velocity for the entire set of threads of $34~\mathrm{km~s^{-1}}$ and a maximum value of $83~\mathrm{km~s^{-1}}$. Blobs move in all directions from horizontal to vertical, with a mean velocity of $51~\mathrm{km~s^{-1}}$ and a maximum of $92~\mathrm{km~s^{-1}}$. 

With our three-dimensional model we have constructed synthetic images including the orientation of the system and LOS integration effects. The synthetic H$\alpha$ images reveal two trapezoidal-shaped prominence segments, each centered at a magnetic dipole and made up primarily of threads. These images clearly show the threads growing and the counterstreaming motion occurring on adjacent threads. We have found that blobs are ubiquitous features in these images, moving along the field lines and falling to the chromosphere. Most of the blobs decelerate before entering the chromosphere. Blobs of the overlying arcade are the coronal rain, with primarily downward motions. Blobs are also produced in the sheared flux tubes associated with the prominence, but their motions are more horizontal than the coronal rain, consistent with observed counterstreaming. We have also computed synthetic images in two EUV channels, 171\AA\ and 211\AA. Most of the predicted 171\AA\ emission comes from the PCTR - the warm plasma between the threads or blobs and the adjacent hot corona. The FC core is brightest in the end view because the concentration of threads and blobs is highest there. In the zone immediately surrounding the core, only blobs are formed so the 171\AA\ emission is collectively dimmer and more variable. In addition, long, bright, transient features occupy most of a tube just before a blob is produced, as the plasma cools through the 171\AA\ passband. In 211\AA, strong emission comes from the FC core in the end-view orientation, with a diffuse halo surrounding the prominence. This indicates that the cool material of the prominence is surrounded by hot material resembling the ``chewy nougats'' seen by Yohkoh \citep{hudson1999} and the observed ``hot prominence shroud'' from recent eclipses \citep{habbal2010}. We have computed the DEM and found very good agreement with the observational DEM of a quiescent prominence \citep{parenti2007}. The computed DEM curve reproduces the observed shape between temperatures $\log\,T = 4.6$ and $5.7$, and the observed position of the minimum at $\log\,T = 5.1$.

The ability of our model to reproduce many observed prominence characteristics lends strong support to the thermal nonequilibrium mechanism for producing the mass of the prominence, as well as the plasma dynamics. The agreement between model and observations also suggests that the fundamental structure of the filament channel is well described by the sheared arcade. Currently, we are extending the realism of our model in three directions, to allow quantitative comparison with prominence and cavity observations. First, we are greatly increasing the number of tubes that sample both the FC and the overlying arcade. Second, in this work have assumed that the footpoint heating is steady in time and is the same on all flux tubes. In an upcoming study we will investigate the effects of a heating function that depends on the key physical parameters, according to the scaling laws summarized in \citet{mandrini2000}, as well as allow for temporally varying heating, consistent with nanoflaring \cite[e.g.,][]{klimchuk2006}. Finally, the emission in EUV coronal lines is significantly affected by Lyman absorption by neutral H and He and by $\mathrm{He}^{+}$ \citep[e.g.,][]{kucera1998}; these important effects will be included in future visualizations of our simulations. The large amplitude oscillations of the condensations in our prominence model will be addressed in an upcoming publication.

\acknowledgments

This work has been supported by the NASA, Heliophysics SR\&T program. M. L. also acknowledges support from the University of Maryland at College Park and the people of CRESST. All of us are grateful to our colleagues on international teams on solar prominences hosted by the International Space Science Institute (ISSI) in Bern, Switzerland, especially team leader N. Labrosse, and acknowledge the support of ISSI, where this work has been presented. Finally, we thank J. A. Klimchuk, T. A. Kucera, S. R. Habbal, and K. K. Reeves for helpful discussions.

\end{document}